\documentclass[pra,aps,amsmath,amssymb,amsfonts,twocolumn,nofootinbib,floatfix]{revtex4}
\usepackage{}
\usepackage{amssymb}
\usepackage{bm,mathrsfs}
\usepackage{graphicx}
\usepackage{epsfig}
\usepackage{amsmath,bbm}
\usepackage{amsfonts,amssymb}
\usepackage{times}
\usepackage{verbatim}
\usepackage[sort&compress]{natbib}
\usepackage{amsmath}
\usepackage{bm}
\usepackage{float}
\usepackage{textgreek}
\usepackage{textcomp}
\allowdisplaybreaks[4]
\usepackage[colorlinks,breaklinks,linkcolor=blue,anchorcolor=blue,citecolor=blue,urlcolor=magenta]{hyperref}

\usepackage{color}
\definecolor{zzz}{rgb}{0.9,0.0,0.4}
\begin{document}

\title{Giant-atom-mediated photon blockade}
% Non-Markovian weak-force amplification with parametric interactions in cavity-optomechanical systems
%Cavity-optomechanical weak-force amplification with parametric interactions and non-Markovian effects}

\author{C. Cui,$^{1}$ W. Y. Hu,$^{1}$  Y. Q. Ji,$^2$ H. T. Cui,$^{3}$ Yan-Hui Zhou,$^{4,}$\footnote{\textcolor{zzz}{Corresponding author: yanhuizhou@126.com}} and H. Z. Shen$^{1,}$\footnote{\textcolor{zzz}{Corresponding author: shenhz458@nenu.edu.cn }}}
\affiliation{$^1$Center for Quantum Sciences and School of Physics, Northeast Normal University, Changchun 130024, China\\
$^2$College of Physics Science and Technology, Bohai University, Jinzhou 121013, China\\
$^3$School of Physics and Optoelectronic Engineering, Ludong University, Yantai 264025, China\\
$^4$Quantum Information Research Center and Jiangxi Province Key Laboratory of Applied Optical Technology, Shangrao Normal University, Shangrao 334001, China}
\date{\today}

\begin{abstract}
Photon blockade is a phenomenon where the presence of system nonlinearity causes the output to consist of single photons, which has been extensively studied in point atom systems, but it is barely explored in giant atom ones. In this paper, we propose giant atom-mediated photon blockade scheme based on two cavities and three cavities systems with driving field applied to the first cavity. We show that simultaneous unconventional photon blockades (UPBs) can not occur in the point atom system (the atom coupling only to the leftmost cavity) due to there always existing a cavity to have a single path. In contrast, the spatially extended nature of giant atom enables coupling to multiple cavities and allows for the introduction of a phase and coupling strength. Consequently, simultaneous UPBs in multiple cavities can be obtained due to the multipath destructive interference. Moreover, by manipulating the detuning, we observe simultaneous conventional photon blockades (CPBs) in multiple cavities. Finally, we study simultaneous two-photon blockades (2PBs) in point atom multiple cavities system.
\end{abstract}

\maketitle
\section{Introduction}
Studies on the generation and manipulation of nonclassical light have become a cornerstone of quantum physics \cite{HJKimble6911977,PRabl0636012011,HZShen0537052023,HZShen0338352013} with photon blockade \cite{Imamoglu14671997} emerging as a major research focus. Photon blockade is a hallmark quantum nonlinear effect where a single photon inhibits the transmission of subsequent photons. It has established a novel paradigm for single-photon devices \cite{YTDeng0337112025,YHZhou0238382015,JTang47052025,HJLi0437072024,XWXu0638532016}, leading to numerous schemes exploiting this effect to generate such sources across diverse platforms. These include cavity quantum electrodynamics (QED) systems \cite{RTrivedi2436022019,KHou0638172019}, superconducting circuits \cite{AJHoffman0536022011}, microresonator waveguide \cite{CCui25762026}, and optomechanical systems \cite{JYSun0437152023,WZZhang0638362015,DYWang0438182019,KBorkje0538332020,DYWang0437052020}. Photon blockade involves two different mechanisms in quantum systems: CPB and UPB.

The first observation of CPB occurred in an optical cavity containing a single trapped atom \cite{KMBirnbaum872005}, which originates from an anharmonicity that is induced by detuning of the system's eigenenergies. This detuning creates a pronounced energy mismatch between single- and two-photon excitations, which prevents a second photon from entering or being emitted simultaneously. CPB mechanism necessitates strong nonlinearities \cite{HZShen0238492014,YHZhou0337132020,SZhao0137122025,QHLiu23004222024,XCGao367962023} obtained through strong coupling to nonlinear quantum elements such as atoms. This effect has been demonstrated in various systems, including cavity QED \cite{MHennrich0536042005,ARidolfo1936022012}, optomechanical architectures \cite{HWang0338062015,XYL0936022015,PKomar0138392013,XNXu0138182015,HXie0638602016,
HXie0138612017,FZou0438372019,YXLiu0321012010}, dynamical blockade \cite{SGhosh0136022019,ZGeng0136022019,GYZhang0237182024,FCavaliere762025}, quantum dots \cite{APFoster1736032019,AFaraon8592008}, three-wave mixing \cite{YRen0537102021}, cavity-coupled two-level systems \cite{ALeBoite0338272016,MRadulaski0118012017,YTGuo0137052022,CSZhao0638382020,ZGLi0437242021}, photon blockades based on topological edge states \cite{wang20255}, and circuit QED \cite{CLang2436012011}. Photon blockade offers potential for a variety of applications such as the interferometers \cite{DGerace2812009,FFratini2436012014}, quantum nonreciprocity \cite{CShang258822019,XYHuang0237032023,XWXu1432020,BJLi752024,YLXiang0437022023}, nonreciprocal CPB \cite{RHuang1536012018,WZhang0237232024,WSXue44242020,YWJing0337072021,XYYao0540042022,YMLiu0637012023,NYuan0535262024,CDGou0437232023}, single-photon transistors \cite{DEChang8072007}, non-Hermitian photon blockade \cite{RHuang21004302022,YLZuo0437152022,JZang1154072015,JHLi0538372015}, multiphoton blockade \cite{FZou0537102020,ZHaider0437022023,GHovsepyan0138392014,CJZhu0638422017,JZLin0538502019,AMiranowicz0138082016,GYZhang250612702,KHou181111409}, and Rydberg blockade \cite{SLSu0440072023,RHZheng0424052023}.

UPB is a phenomenon that enables single-photon emission through the manipulation of destructive interference pathways with weak nonlinearities \cite{TCHLiew1836012010,MBamba0218022011,HFlayac0338632013,DGerace0318032014,OKyriienko1974022020,HJabri0237042022,BijitaSarma0138262018,HFlayac0538102017,EZCasalengua19002792020}, which bypasses CPB's strong-nonlinearity requirement \cite{ZGLu0136022025,ICarusotto2992013,YHZhou4722019,GCWang5832017,YHZhou29352017,YHZhou0640092022,HXZheng2236012011,HZShen0638082015,HYSun36402019,YWang2404022021,SYLi24003742025,HZShen0355032018}. UPB has been experimentally observed in the microwave domain using coupled superconducting resonators \cite{ CVaneph0436022018} and quantum dot cavity QED systems \cite{HJSnijders0436012018}. It is predicted to occur in numerous quantum systems such as nonlinear photonic molecules \cite{XWXu0438222014,XWXu0338092014}, quantum dots \cite{WZhang0438322014,XYLiang0537132020,AMajumdar1836012012,JTang0440652019,JHLi0538012018,TFFang1554172011}, second- or third-order nonlinearity \cite{DRoberts0210222020,HFlayac0138152016,HZShen328352015,YHZhou173322016,
OKyriienko0638052014,MLi0436012022,YCLi0437022024,ZHLiu0637052024,XFQiao0537022024}, cavity optomechanics \cite{YQu0438232019,LLZheng0138042019}, gain cavities \cite{YHZhou0438192018}, excited polaritons \cite{JCLopez1964022015}, non-Markovian effects \cite{HZShen0238562018,HZShen0437142024}, Gaussian squeezing \cite{BSarma0538272017,MALemonde0638242014}, and nonreciprocal UPB \cite{BJLi6302019,HZShen0138262020,JWang640032021,TZLuan23500212023,HZShen23500292023,JXYang250510255}. Moreover, recent study proposed the enhanced mechanisms for universal photon blockade \cite{YHZhou1836012025,YHZhou0337222026}.

Giant atoms are generally formed when artificial atoms are coupled either to propagating fields characterized by wavelengths much smaller than the atom scale \cite{AFKockum0138372014,LDu2236022022,ANoguchi1805052017,KJSatzinger6612018,BAMoores2277012018,ANBolgar2236032018,ABienfait3682019,
GAndersson2404022020} or to a waveguide with a meandering path at well-separated coupling points \cite{AMVadiraj0237102021,ZMGao0537062024, ZMGao0137162024,MZWeng0237212024,YTCheno0637102024,WZJia94952024,WJGu0237202024,YYYan0133012024,XLYin0637032022}. They originate from their breakdown of the dipole approximation \cite{AFKockum1252021,ASoro0237122022,ZYLi0237122024,SJSun10340}, as their spatial extent becomes comparable to the wavelength of interacting electromagnetic fields. This allows them to couple to the field at multiple discrete points, which in turn gives them distinct advantages in studying photon blockade phenomena \cite{XWang0337022024,CMZheng0430302023,KJMa0251092025}. Moreover, experimental investigations of giant atoms have witnessed significant progress in non-Markovian dynamics \cite{GAndersson11232019}, manipulation of energy level structures \cite{BKannan7752020}, and coupling techniques with waveguides such as surface acoustic waves \cite{RManenti9752017,MVGustafsson2072014}. In giant atom systems, novel phenomena arising from quantum interference effects between multiple coupling points have been predicted. These include decoherence-free interaction \cite{AFKockum1404042018,DCilluffo0430702020,ASoro0137102023}, electromagnetically induced transparency \cite{YTZhu0437102022}, non-exponential relaxation dynamics \cite{LZGuo0538212017,SJGuo0337062020}, formation of bound states \cite{LZGuo0430142020,KHLim0237162023,WZhao0538552020,XWang0436022021,HXiao802022,CVega0535222021,HZShen315042019}, and other derivative effects \cite{DDNoachtar0137022022,ACSantos0536012023,
QYQiu2242122023,ZQWang75802022,XWang0437032022,STerradasBrianso0637172022,SARegidor0337192023,ESanchezBurillo0137092020,XWang0132792024,XLYin0237282023}. The giant atom platform thus provides an effective approach for photon manipulation \cite{WJGu0537182023,YPPeng0437092023,QYCai0337102021,SLFeng0637122021,XLYin0137152022}, particularly enabling nonreciprocal photon propagation \cite{HWYu0137202021,JZhou0637032023,YTChen2152022}.

In this paper, we construct giant atom two cavities and three cavities systems to study simultaneous photon blockades in multiple cavities, where driving field mediates the first cavity. In point atom systems where the atom is coupled only to the leftmost cavity, the rightmost cavity lacks destructive interference paths, which prevents simultaneous generation of UPBs in multiple cavities. In contrast, the giant atom can couple to multiple cavities at two different coupling points, creating an additional destructive interference path, which gives the optimal conditions for producting simultaneous UPBs in multiple cavities. Moreover, we derive the eigenfrequencies and optimal detunings for single- and two-excitation subspaces. Due to energy-level anharmonicity, CPB occurs simultaneously in multiple cavities. Simultaneous 2PBs are also studied in point atom multiple cavities system.

The paper is organized as follows. In Sec.~\ref{er}, we establish a model of giant atom coupled to two cavities and three cavities systems. In Sec.~\ref{san}, we study simultaneous CPBs and simultaneous UPBs in giant atom and two cavities system. In Sec.~\ref{si}, simultaneous UPBs, simultaneous CPBs, and simultaneous 2PBs for point atom and two cavities system are presented. In Sec.~\ref{wu}, we investigate influences of the giant atom effects on simultaneous CPBs and simultaneous UPBs in three cavities. Sec.~\ref{liu} is devoted to conclusions.

\section{Model Hamiltonian}
\label{er}

\begin{figure}[t]
\centering{
\includegraphics[width=8.5cm,  height=4.4cm,  clip]{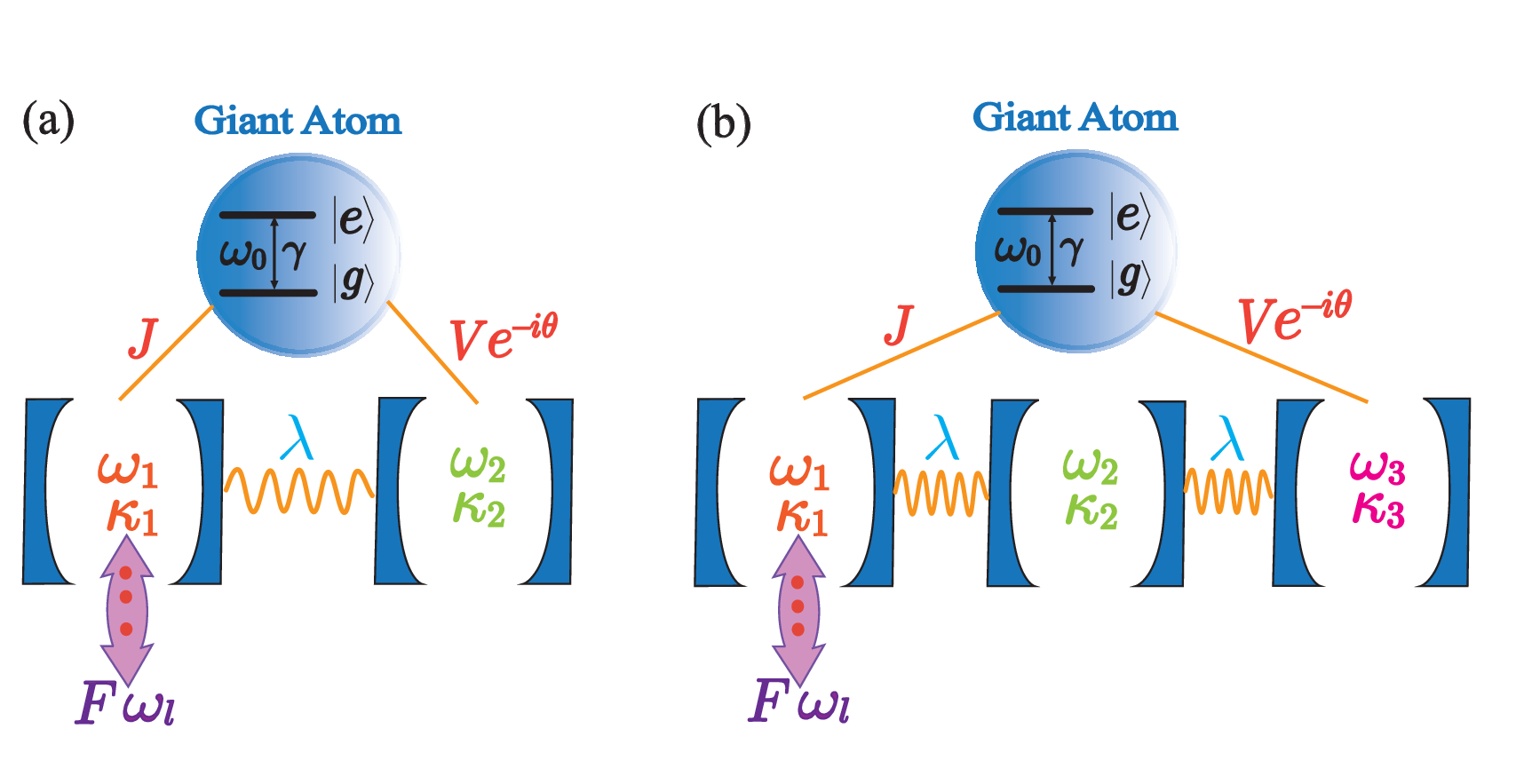}}
\caption{Schematic of a two-level giant atom (transition frequency $\omega_0$, dissipation $\gamma$) coupled to (a) two cavities and (b) three cavities, where the intercavity coupling coefficient is $\lambda$. The atom couples to the first cavity (frequency $\omega_1$, dissipation $\kappa_1$) with real coupling coefficient $J$ and (a) to the second cavity (frequency $\omega_2$, dissipation $\kappa_2$) with complex coupling coefficient $V e^{-i\theta}$ (strength $V$, phase $\theta$), (b) to the third cavity (frequency $\omega_3$, dissipation $\kappa_3$) with complex coupling coefficient $V e^{-i\theta}$, where the second cavity's frequency and dissipation are $\omega_2$ and $\kappa_2$, respectively. The first cavity is mediated via driving field (strength $F$, frequency $\omega_l$).}
\label{setup}
\end{figure}

To investigate photon blockades mediated by a giant atom, we employ the giant atom coupled to two cavities in Fig.~\ref{setup}(a) and three cavities in Fig.~\ref{setup}(b), where cavity \( a_1 \) is manipulated by driving field. The corresponding system Hamiltonians in Fig.~\ref{setup} (\( \hbar \equiv 1 \)) read
%\centering\scalebox{0.43}{\includegraphics{setupv2-01.eps}}
% 图1
\begin{equation}
\begin{aligned}
\hat{\mathcal{H}}_2 &= \omega_1 \hat{a}_1^\dag \hat{a}_1 + \omega_2 \hat{a}_2^\dag \hat{a}_2 + \omega_0 \sigma^+ \sigma + J( \hat{a}_1\sigma^+ + \hat{a}_1^\dag \sigma)  \\
&\quad + V(e^{-i\theta} \hat{a}_2 \sigma^+ + e^{i\theta} \hat{a}_2^\dag \sigma) + \lambda (\hat{a}_1 \hat{a}_{2}^\dag + \hat{a}_{2} \hat{a}_1^\dag)  \\
&\quad + F(e^{i\omega_l t} \hat{a}_1+ e^{-i\omega_l t} \hat{a}_1^\dag) , \\
\hat{\mathcal{H}}_3 &= \omega_1 \hat{a}_1^\dag \hat{a}_1 + \omega_2 \hat{a}_2^\dag \hat{a}_2 + \omega_3 \hat{a}_3^\dag \hat{a}_3 + \omega_0 \sigma^+ \sigma + J( \hat{a}_1\sigma^+ \\
&\quad  + \hat{a}_1^\dag \sigma) + V(e^{-i\theta} \hat{a}_3 \sigma^+ + e^{i\theta} \hat{a}_3^\dag \sigma) + \lambda (\hat{a}_1 \hat{a}_{2}^\dag + \hat{a}_{2} \hat{a}_1^\dag) \\
&\quad  + \lambda (\hat{a}_2 \hat{a}_{3}^\dag + \hat{a}_{3} \hat{a}_2^\dag)+ F(e^{i\omega_l t} \hat{a}_1  + e^{-i\omega_l t} \hat{a}_1^\dag), \label{Hamiltonian3}
\end{aligned}
\end{equation}
where $\omega_j$ and $\omega_0$ respectively represent the frequencies of two (or three) cavities (annihilation operator $\hat a_j$, $j = 1,2,3$) and giant atom (lowering operator $\sigma$). $J$ denotes the real coupling coefficient between giant atom and first cavity, while $V e^{-i\theta}$ with strength $V$ and phase $\theta$ represents complex coupling coefficient between giant atom and end cavity (second or third cavity in this case). $\lambda$ describes the coupling coefficient between different cavities. $F$ and $\omega_l$ respectively denote the strength and frequency of driving field acting on the first cavity. The two Hamiltonians with a rotating frame in Eq.~(\ref{Hamiltonian3}) respectively become
\begin{align}
\hat H_2 &= \Delta_1 \hat a_1^\dag \hat a_1 + \Delta_2 \hat a_2^\dag \hat a_2 + \Delta_0 \sigma^+ \sigma + J(\hat a_1 \sigma^+ + \hat a_1^\dag \sigma) \nonumber\\
&\quad + V(e^{-i\theta} \hat a_2 \sigma^+ + e^{i\theta} \hat a_2^\dag \sigma) + \lambda(\hat a_1 \hat a_2^\dag + \hat a_2 \hat a_1^\dag) \nonumber\\
&\quad + F(\hat a_1 + \hat a_1^\dag), \label{H2} \\
\hat H_3 &= \Delta_1 \hat a_1^\dag \hat a_1 + \Delta_2 \hat a_2^\dag \hat a_2 + \Delta_3 \hat a_3^\dag \hat a_3 + \Delta_0 \sigma^+ \sigma \nonumber\\
&\quad + J(\hat a_1 \sigma^+ + \hat a_1^\dag \sigma) + V(e^{-i\theta} \hat a_3 \sigma^+ + e^{i\theta} \hat a_3^\dag \sigma) \nonumber\\
&\quad + \lambda (\hat a_1 \hat a_2^\dag + \hat a_2 \hat a_1^\dag) + \lambda (\hat a_2 \hat a_3^\dag + \hat a_3 \hat a_2^\dag) + F(\hat a_1 + \hat a_1^\dag), \label{H3}
\end{align}
where ${\Delta _1} = {\omega _1} - {\omega _l}$, ${\Delta _2} = {\omega _2} - {\omega _l}$ (${\Delta _3} = {\omega _3} - {\omega _l}$), and ${\Delta _0} = {\omega _0} - {\omega _l}$ represent the detunings of two (three) cavities and giant atom from the driving field, respectively. Considering dissipations, the evolutions for two cavities and three cavities systems are respectively governed by
\begin{align}
   \dot \rho_2 = - i[\hat H_2,\rho_2 ] +  \sum_{j=1}^{2} \kappa_{{j}} {{\cal L}(a_j)}\rho_2 + \gamma {{\cal L}(\sigma)}\rho_2 , \label{rou2}
\end{align}
and
\begin{align}
   \dot \rho_3 = - i[\hat H_3,\rho_3 ] +  \sum_{j=1}^{3} \kappa_{{j}} {{\cal L}(a_j)}\rho_3 + \gamma {{\cal L}(\sigma)}\rho_3 ,
   \label{rou3}
\end{align}
where ${{\cal L}}(\hat{o})\rho  = \hat{o}\rho \hat{o}^\dagger - \frac{1}{2}\{ \hat{o}^\dagger \hat{o}, \rho \} $ is the Lindblad superoperator for operator \( \hat{o} \). $\hat H_2 $ and $\hat H_3 $ are respectively given by Eqs.~(\ref{H2}) and (\ref{H3}). \( \rho_2 \) and \( \rho_3 \) are the density matrices of two cavities and three cavities systems. \( \kappa_{j} \) and \( \gamma \) represent the dissipations of the $j$-th cavity and giant atom, respectively. Hereafter, we set $\kappa_1=\kappa_2=\kappa_3=\gamma \equiv\kappa$ for simplification. We focus on the steady-state second-order correlation function
\begin{eqnarray}
g_{ o}^{(2)}(0) = \frac{{\left\langle {{{\hat o}^\dag }{{\hat o}^\dag }\hat o\hat o} \right\rangle }}{{{{\left\langle {{{\hat o}^\dag }\hat o} \right\rangle }^2}}},
   \label{go20}
\end{eqnarray}
serving as a measure for identifying single-photon blockade. Sub-Poissonian and super-Poissonian statistics properties are indicated by \(g^{(2)}(0)<1\) and \(g^{(2)}(0)>1\), which correspond to photon antibunching and bunching effects, respectively.
\section{Simultaneous Photon Blockades in two cavities with giant atom}
\label{san}
\subsection{CPBs occur simultaneously in two cavities}
\label{sanA}
In this section, we consider giant atom coupled to two cavities system determined by Eq.~(\ref{H2}). The states of the system can be described by the basis $|mng(e)\rangle$, where $|m\rangle$ and $|n\rangle$ respectively denote the photon number states of cavities $a_1$ and $a_2$, while $|g\rangle$ and $|e\rangle$ represent the atom ground and excited states. Assuming $\omega_1 = \omega_2 = \omega_0 \equiv \omega'$ (or setting ${\Delta _1} = {\Delta _2} = {\Delta _0} \equiv \Delta $), $\theta = 0$, and $\lambda = J \equiv V$, we choose the states $\lvert 10g \rangle$, $\lvert 01g \rangle$, and $\lvert 00e \rangle$ to extend Hamiltonian (\ref{H2}) without the driving terms as
\begin{equation}
\hat H'_2 = \left( {\begin{array}{*{20}{c}}
{{\omega'}}&V &V\\
V &{{\omega'}}&{V}\\
V&{V}&{{\omega'}}
\end{array}} \right),
\label{H2CPB1}
\end{equation}
which yields three eigenfrequencies
\begin{equation}
\begin{aligned}
\omega _{1,2}^{(1)} = \omega'  - V, \quad \omega _3^{(1)} = \omega'  + 2V,
\label{2omega1}
\end{aligned}
\end{equation}
where the superscript (1) describes the single-excitation subspace. Setting the driving frequency $\omega_l$ resonant with the eigenfrequencies in Eq. (\ref{2omega1}), we obtain the optimal detunings
% 图2
\begin{figure}[t]
\centering\scalebox{0.28}{\includegraphics{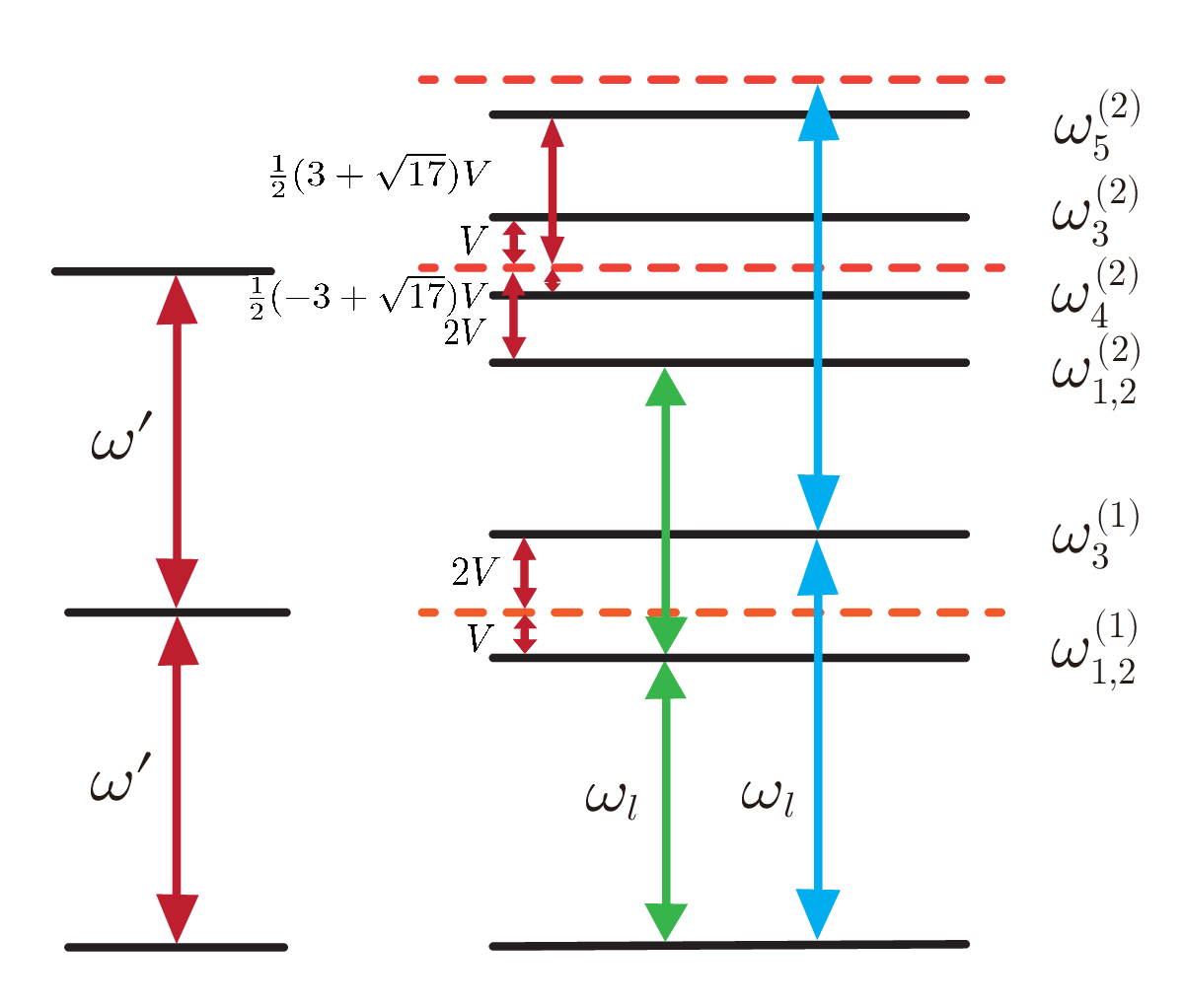}}
\caption{Level diagram of CPB mechanism for giant atom and two cavities system. The eigenfrequencies are indicated by horizontal lines. The frequency \(\omega'\) on the left denotes the bare frequency of the uncoupled system, serving as a common reference for detuning \(\Delta\). The blue arrow corresponds to $\Delta_{\text{opt}} = -2V$ obtained by Eq.~(\ref{2Delta1}), while the green arrow marks $\Delta_{\text{opt}} = V$ given by Eq. (\ref{2Delta2}). Here we only show several main pathways of two-photon transitions.}
\label{CPBlevel0}
\end{figure}
\begin{align}
\Delta_{\text{opt}}  = V\quad\text{or}\quad\Delta_{\text{opt}}= - 2V.
\label{2Delta1}
\end{align}

In the two-excitation subspace, the matrix form of the Hamiltonian (\ref{H2}) without the driving terms reads
\begin{equation}
\hat H''_2 =
\begin{pmatrix}
2\omega' & 0 & \sqrt{2}V & \sqrt{2}V & 0 \\
0 & 2\omega' & \sqrt{2}V & 0 & \sqrt{2}V \\
\sqrt{2}V & \sqrt{2}V & 2\omega' & V & V \\
\sqrt{2}V & 0 & V & 2\omega' & V \\
0 & \sqrt{2}V & V & V & 2\omega'
\end{pmatrix},
\label{H2CPB2}
\end{equation}
with the eigenfrequencies
% 图3
\begin{figure}[t]
\centering\scalebox{0.40}{\includegraphics{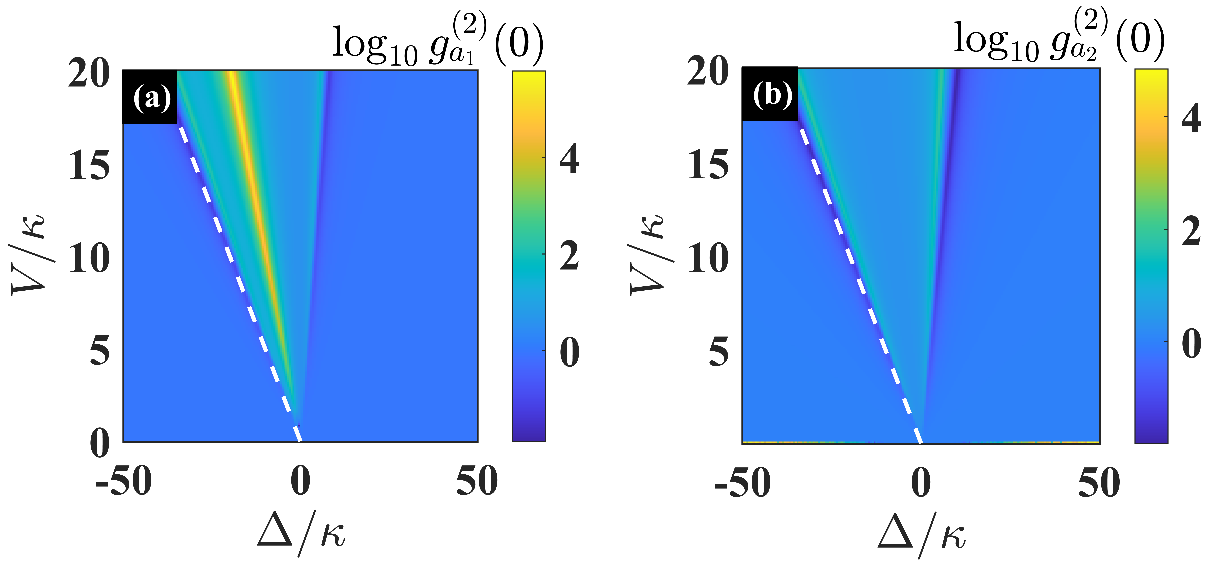}}
\caption{Second-order correlation function (solved by Eq.~(\ref{rou2})) with giant atom and two cavities system on a logarithmic scale \(\log_{10}g^{(2)}_{a_1}(0)\) and \(\log_{10}g^{(2)}_{a_2}(0)\) as a function of detuning \( \Delta \) and coupling strength \( V \) with \( a_1 \) in (a) and \( a_2 \) in (b). CPBs occur simultaneously in two cavities at the same location at \(\Delta_{\text{opt}} = -2V\) in Eq.~(\ref{2Delta1}) (corresponding to two white dashed lines) for (a)(b). The other parameters chosen are \( F = 0.01\kappa \) and \(\theta=0\).}
\label{CPB}
\end{figure}
\begin{equation}
\begin{aligned}
\omega _{1,2}^{(2)} &= 2{\omega'} - 2V,\\
\omega _3^{(2)} &= 2{\omega'} + V,\\
\omega _4^{(2)} &= 2{\omega'} - \frac{1}{2}( { - 3 + \sqrt {17} })V ,\\
\omega _5^{(2)} &= 2{\omega'} + \frac{1}{2}( {3 + \sqrt {17} })V ,
\label{2omega2}
\end{aligned}
\end{equation}
where the superscript (2) denotes the two-excitation subspace. The two-photon resonance with 2$\omega_l$ equaling eigenfrequencies in Eq.~(\ref{2omega2}) gives optimal detunings
\begin{equation}
\begin{aligned}
\Delta_{\text{opt}}  =& \,V,\,\,\,\,\,\,\,\,\,\,\Delta_{\text{opt}}  = -\frac{1}{2}V,\,\,\,\,\,\,\,\,\,\,\,\Delta_{\text{opt}}  = \frac{1}{4}( { - 3 + \sqrt {17} } )V,\\
\Delta_{\text{opt}}=& - \frac{1}{4}( {  3 + \sqrt {17} } )V.
\label{2Delta2}
\end{aligned}
\end{equation}

Figure \ref{CPBlevel0} provides a detailed analysis of CPB, showing the eigenfrequencies as black horizontal lines to illustrate the conditions of CPB generation derived from Eqs.~(\ref{2omega1}) and (\ref{2omega2}). In Fig.~\ref{CPBlevel0}, when the system is driven by an incident laser of frequency $\omega_l$, an incident photon with the same frequency as the cavity $\omega_{3}^{(1)}$ excite two cavities from the vacuum state $|00\rangle$ to the first excited states $|10\rangle$ and $|01\rangle$. Once each of two cavities contains a photon simultaneously, the second photon (transition $\omega_{3}^{(1)} \rightarrow \omega_{5}^{(2)}$) will be blocked owing to the large detuning of $\left(4 - \frac{1}{2}(3 + \sqrt{17})\right) V\approx 0.4384V$. In Fig.~\ref{CPBlevel0}, the blue solid arrow indicates CPB at the optimal detuning condition $\Delta_{\text{opt}} = -2V$ in Eq.~(\ref{2Delta1}), which leads to the suppression of two-photon transitions.

Figure~\ref{CPB} shows logarithmic plots of the second-order correlation function \( g^{(2)}(0) \) for cavities \( a_1 \) and \( a_2 \) as a function of detuning \( \Delta \) and coupling strength \( V \). In Fig.~\ref{CPB}(a)(b), the strong photon antibunching ($g^{(2)}(0) \ll 1$) is simultaneously observed along $\Delta_{\text{opt}}= -2V$ (see two white dashed lines) from Eq.~(\ref{2Delta1}), which confirms the presence of simultaneous CPBs in two cavities. This phenomenon arises from the anharmonicity of the system's energy level structure in Fig.~\ref{CPBlevel0}.

\subsection{UPBs occur simultaneously in two cavities}
\label{sanB}
In this section, we derive the optimal condition for getting simultaneous UPBs in two cavities by setting ${\Delta _1} = {\Delta _2} = {\Delta _0} \equiv \Delta $. The effective Hamiltonian from Eq.~(\ref{rou2}) is
\begin{equation}
{{\hat H}_{\text{eff}}} = \hat H_2 - \frac{i}{2}{\kappa}\hat a_1^\dag {{\hat a}_1} - \frac{i}{2}{\kappa}\hat a_2^\dag {{\hat a}_2} - \frac{i}{2}\kappa {\sigma ^ + }\sigma.
   \label{H2eff}
\end{equation}
The atom is initially prepared in the ground state and cavities are in the vacuum state. Under the weak driving condition, the state of the system is
\begin{align}
\left| \psi  \right\rangle  &= {C_{00g}}\left| {00g} \right\rangle  + {C_{10g}}\left| {10g} \right\rangle  + {C_{01g}}\left| {01g} \right\rangle \nonumber\\
&\quad+ {C_{00e}}\left| {00e} \right\rangle  + {C_{20g}}\left| {20g} \right\rangle  + {C_{02g}}\left| {02g} \right\rangle \nonumber\\
&\quad+ {C_{11g}}\left| {11g} \right\rangle  + {C_{10e}}\left| {10e} \right\rangle  + {C_{01e}}\left| {01e} \right\rangle,
   \label{2pusai}
\end{align}
where \( C_{mng} \) and \( C_{mne} \) are the steady-state probability amplitudes for the states \(  |mng\rangle \) and \(  |mne\rangle\), respectively. We derive \( C_{20g} \) and \( C_{02g} \) in Appendix \ref{A} under the weak driving condition with $C_{00g} \simeq 1$ $\gg$ $C_{10g}$, $C_{01g}$, $C_{00e}$ $\gg$ $C_{20g}$, $C_{02g}$, $C_{11g}$, $C_{10e}$, $C_{01e}$. Based on Eqs.~(\ref{go20}) and (\ref{2pusai}), the weak driving approximation gives
\begin{align}
g_{{ a_1}}^{(2)}(0) \simeq \frac{{2{{\left| {{C_{20g}}} \right|}^2}}}{{{{\left| {{C_{10g}}} \right|}^4}}},\quad g_{{ a_2}}^{(2)}(0) \simeq \frac{{2{{\left| {{C_{02g}}} \right|}^2}}}{{{{\left| {{C_{01g}}} \right|}^4}}}.
\label{2g20}
\end{align}
Simultaneous UPBs in two cavities require $g_{a_1}^{(2)}(0) = g_{a_2}^{(2)}(0) = 0$ in Eq.~(\ref{2g20}) with optimal condition
% 图4
\begin{figure}[t]
\centering{
\includegraphics[width=8.6cm,  height=6.3cm,  clip]{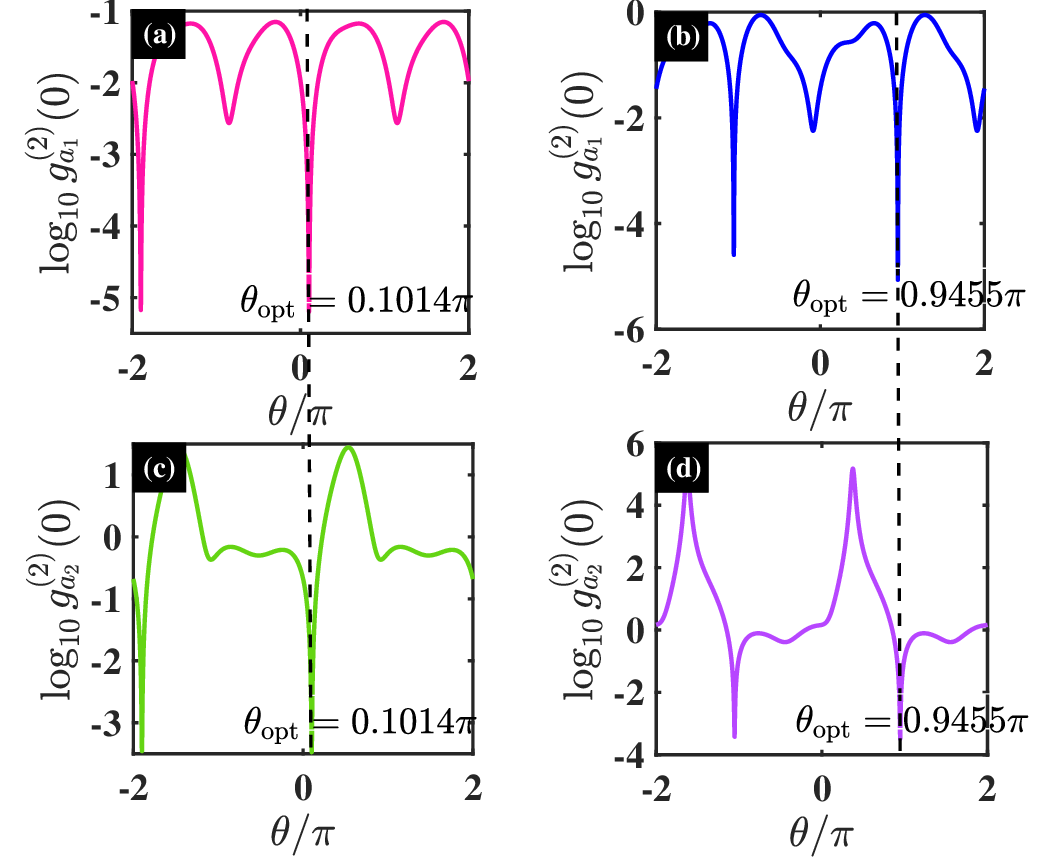}}
\caption{\(\log_{10}g^{(2)}_{o}(0)\) (solved by Eq.~(\ref{rou2})) with giant atom and two cavities system as a function of phase $\theta$ with $a_1$ in (a)(b) and $a_2$ in (c)(d). The black dashed lines denote the optimal phase $\theta_{\text{opt}}$ given by Eq.~(\ref{2optimal}). The parameters chosen are (a)(c) $\lambda = 4\kappa$, $J_{\text{opt}} = -0.7379\kappa$, $V_{\text{opt}} = 1.4637\kappa$, \(F=0.01\kappa\), and $\Delta_{\text{opt}} = 0.0805\kappa$; (b)(d) $\lambda = 5.9693\kappa$, $J_{\text{opt}} = -1.4993\kappa$, $V_{\text{opt}} = 2\kappa$, \(F=0.015\kappa\), and $\Delta_{\text{opt}} = -0.2095\kappa$. UPBs occur simultaneously in two cavities at the same location at \(\theta_{\text{opt}} = 0.1014\pi\) (with period 2$\pi$) for (a)(c), also at \(\theta_{\text{opt}} = 0.9455\pi\) (with period 2$\pi$) for (b)(d).}
\label{one}
\end{figure}
% 图5
\begin{figure}[t]
\centering{
\includegraphics[width=8.7cm,  height=6.5cm,  clip]{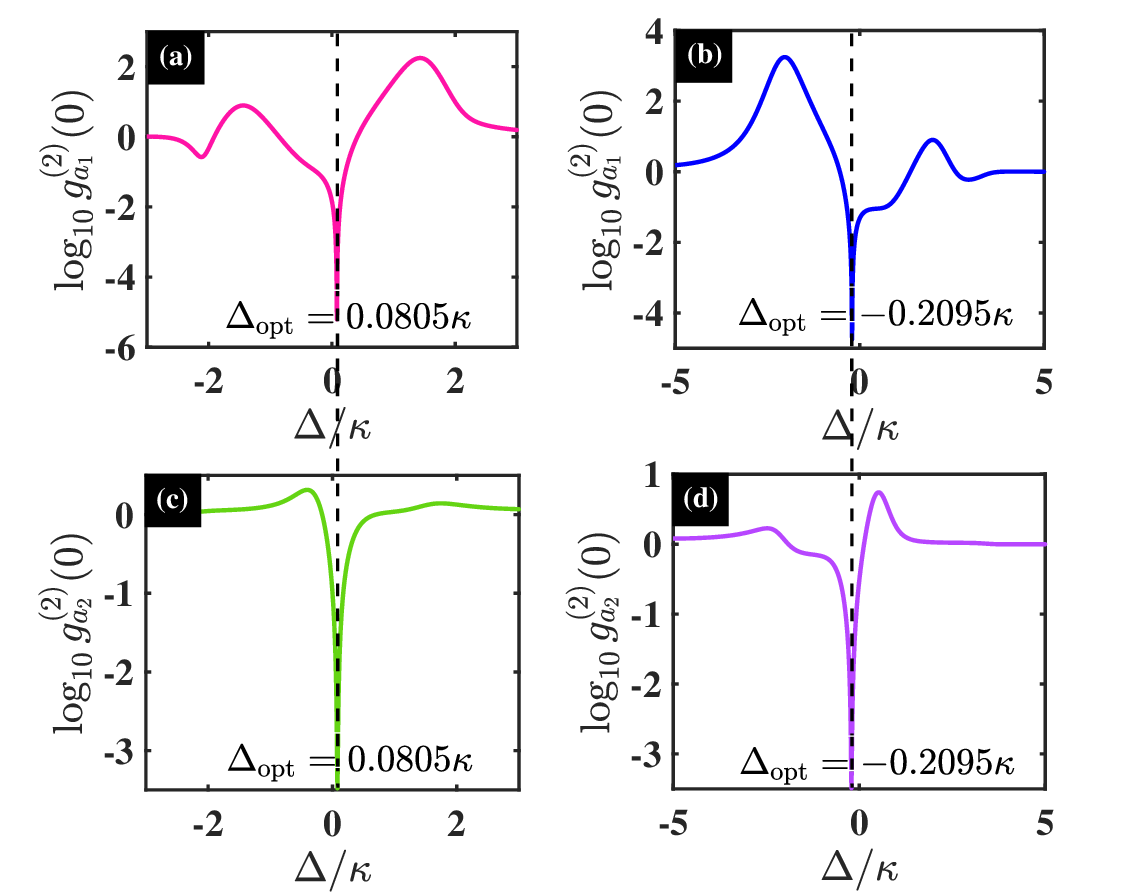}}
\caption{\(\log_{10}g^{(2)}_{o}(0)\) (solved by Eq.~(\ref{rou2})) with giant atom and two cavities system as a function of detuning \(\Delta\) with \(a_1\) in (a)(b) and \(a_2\) in (c)(d) at optimal phases determined by Eq.~(\ref{2optimal}). The parameters chosen are (a)(c) \(\theta_{\text{opt}} = 0.1014\pi\); (b)(d) \(\theta_{\text{opt}} = 0.9455\pi\). The black dashed lines correspond to the optimal detuning $\Delta_{\text{opt}}$ given by Eq.~(\ref{2optimal}). UPBs occur simultaneously in two cavities at the same location at $\Delta_{\text{opt}} = 0.0805\kappa$ for (a)(c), also at $\Delta_{\text{opt}} = -0.2095\kappa$ for (b)(d). The other parameters chosen are the same as those in Fig.~\ref{one}.}
\label{oneDelta}
\end{figure}
% 图6
\begin{figure}[t]
\centering\scalebox{0.42}{\includegraphics{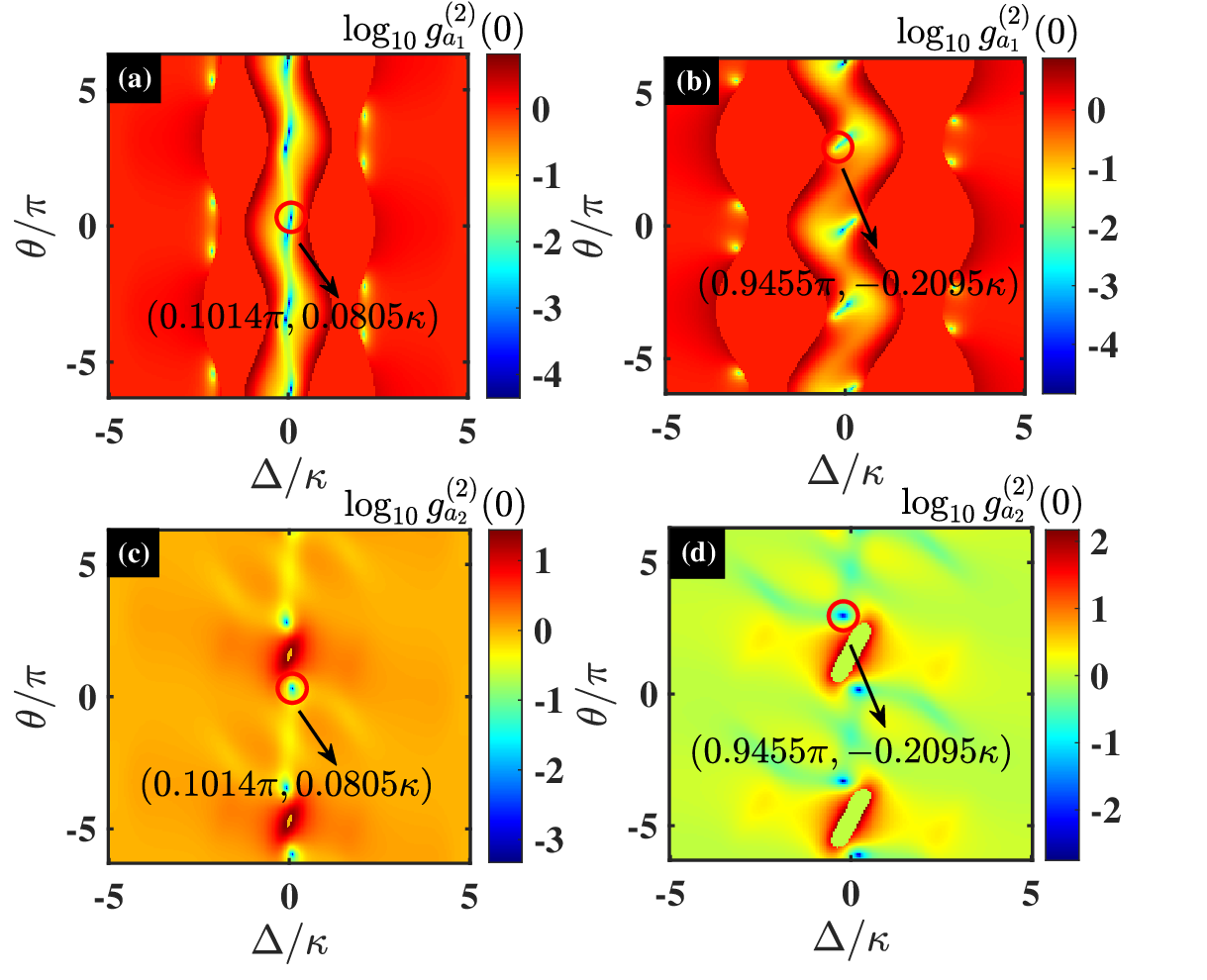}}
\caption{\(\log_{10}g^{(2)}_{o}(0)\) (solved by Eq.~(\ref{rou2})) with giant atom and two cavities system as a function of detuning \(\Delta\) and phase \(\theta\) with \(a_1\) in (a)(b) and \(a_2\) in (c)(d). The red circles mark the optimal points. UPBs occur simultaneously in two cavities at the same location at \(\theta_{\text{opt}}=0.1014\pi\) (with period 2$\pi$) and \(\Delta_{\text{opt}}=0.0805\kappa\) for (a)(c), also at \(\theta_{\text{opt}} = 0.9455\pi\) (with period 2$\pi$) and \(\Delta_{\text{opt}} = -0.2095\kappa\) for (b)(d) determined by Eq.~(\ref{2optimal}). The other parameters chosen are the same as those in Fig.~\ref{one}.}
\label{oneDeltatheta}
\end{figure}
% 图7
\begin{figure}[t]
\centering\scalebox{0.48}{\includegraphics{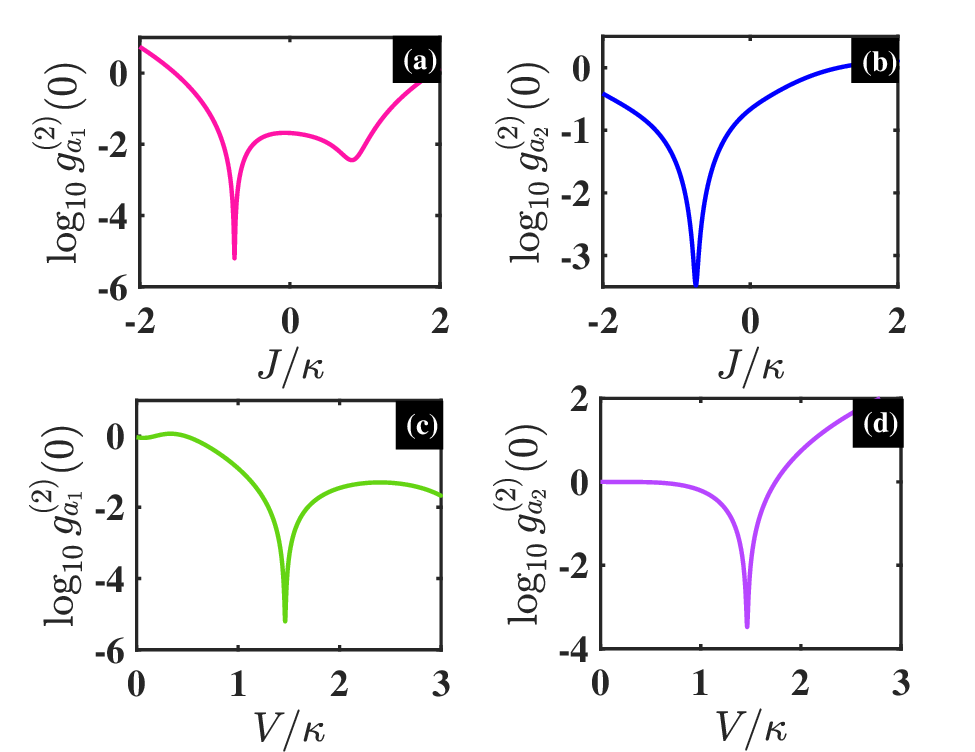}}
\caption{\(\log_{10}g^{(2)}_{o}(0)\) (solved by Eq.~(\ref{rou2})) with giant atom and two cavities system as functions of coupling strength \(J\) and coupling strength \(V\) with \(a_1\) in (a)(c) and \(a_2\) in (b)(d). UPBs occur simultaneously in two cavities at the same location at \(J_{\text{opt}}=-0.7379\kappa\) for (a)(b), also at \(V_{\text{opt}}=1.4637\kappa\) for (c)(d) determined by Eq.~(\ref{2optimal}). The other parameters chosen are the same as those in Fig.~\ref{one}(a)(c).}
\label{upb_F_V_2}
\end{figure}
\begin{align}
\operatorname{Re}(A) = 0,\,\,\, \operatorname{Im}(A) = 0,\,\,\,\operatorname{Re}(C) = 0, \,\,\,\operatorname{Im}(C) = 0,
\label{2optimal}
\end{align}
where \(A\) and \(C\) are provided in Appendix \ref{A}. Using Eq.~(\ref{2optimal}), we plot \( g^{(2)}(0) \) as a function of phase \( \theta \) in Fig.~\ref{one}. We select the optimal parameters \(\lambda_{\text{opt}}\), \(J_{\text{opt}}\), \(V_{\text{opt}}\), \(\Delta_{\text{opt}}\), and \(\theta_{\text{opt}}\) determined by Eq.~(\ref{2optimal}). The black dashed line common to Fig.~\ref{one}(a)(c) indicates simultaneous occurrence of a strong antibunching effect in two cavities at \(\theta_{\text{opt}} = 0.1014\pi\). Moreover, we find that the strong photon antibunching effect recurs with period 2$\pi$, corresponding to \(\theta_{\text{opt}} = -1.8986\pi\) in Fig.~\ref{one}(a)(c). By increasing the intercavity coupling strength to \(\lambda = 5.9693\kappa\), simultaneous UPBs are again observed at \(\theta_{\text{opt}} = 0.9455\pi\), which is illustrated by the black dashed line in Fig.~\ref{one}(b)(d).

We plot \( g^{(2)}(0) \) as a function of detuning \(\Delta\) in Fig.~\ref{oneDelta}. With the same optimal parameters as in Fig.~\ref{one}, the black dashed line spanning Fig.~\ref{oneDelta}(a)(c) shows simultaneous UPBs in two cavities at $\Delta_{\text{opt}} = 0.0805\kappa$ for $\lambda = 4\kappa$ and $\theta_{\text{opt}} = 0.1014\pi$ determined by Eq.~(\ref{2optimal}), which is consistent with the detuning parameter value used in Fig.~\ref{one}. When \(\lambda\) is added to $5.9693\kappa$, the same effect is observed at \(\Delta_{\text{opt}} = -0.2095\kappa\) given by Eq.~(\ref{2optimal}) in Fig.~\ref{oneDelta}(b)(d). In Fig.~\ref{oneDeltatheta} a three-dimensional plot in \( g^{(2)}(0) \) as a function of phase \(\theta\) and detuning \(\Delta\) reveals the control of these two parameters in producing simultaneous UPBs. In Fig.~\ref{oneDeltatheta}(a)(c), the optimal points are marked by red circles, exhibiting simultaneous UPBs in two cavities and corresponding to the optimal parameters identified in Fig.~\ref{one}(a)(c) and Fig.~\ref{oneDelta}(a)(c). Similarly, Fig.~\ref{oneDeltatheta}(b)(d) demonstrates consistent behavior, where \(g^{(2)}(0)\) reaches minimum values at optimal parameters marked by red circles.
% 图8
\begin{figure}[t]
\centering\scalebox{0.23}{\includegraphics{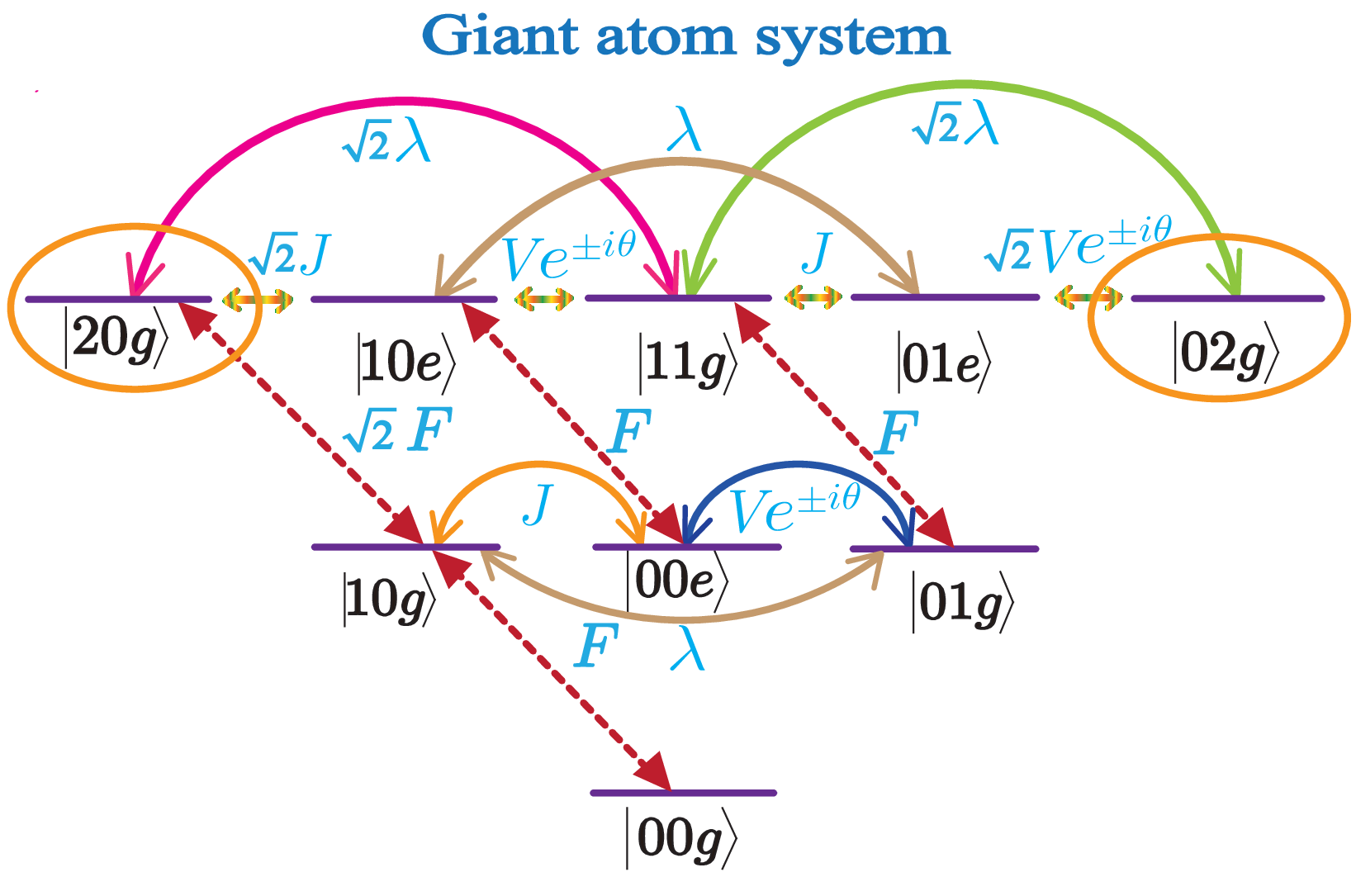}}
\caption{Level diagram of giant atom coupled to two cavities system. It depicts the zero-, one-, and two-photon states and transition paths responsible for the destructive interference that induces strong photon antibunching. States are labeled as $|mn\alpha\rangle$, where $|m\rangle$ ($|n\rangle$) denotes photon number state in cavity $a_1$ ($a_2$), while $|\alpha\rangle$ ($\alpha=g,e$) indicates the atom state.}
\label{level}
\end{figure}
% 图9
\begin{figure}[h]
\centering{\includegraphics[width=8.8cm,  height=6.3cm,  clip]{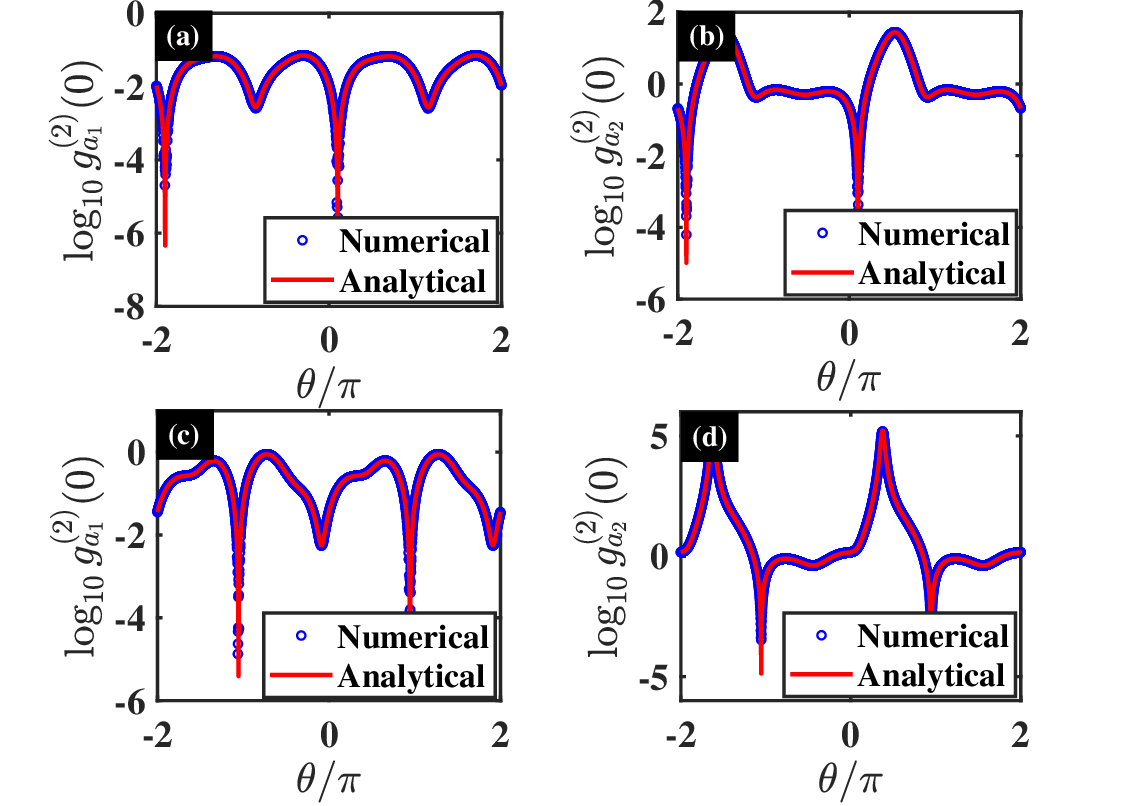}}
\caption{\(\log_{10}g^{(2)}_{o}(0)\) (solved by Eq.~(\ref{rou2})) with giant atom and two cavities system as a function of phase $\theta$ with $a_1$ in (a)(c) and $a_2$ in (b)(d). The red solid lines represent results based on the analytical expression (\ref{2g20}), while the blue circles show numerical simulations from the master equation (\ref{rou2}). The parameters chosen are the same as those in Fig.~\ref{one}.}
\label{shuzhijiexi}
\end{figure}

Moreover, we respectively investigate the influences of coupling strength \( J \) and coupling strength \( V \) on simultaneous UPBs in two cavities in Fig.~\ref{upb_F_V_2}. In Fig.~\ref{upb_F_V_2}(a)(b), \( g^{(2)}(0) \) of two cavities reaches minima at optimal coupling strength \(J_{\text{opt}} = -0.7379\kappa\) given by Eq.~(\ref{2optimal}), indicating the realization of simultaneous UPBs in two cavities. Similarly, simultaneous UPBs occur in two cavities at optimal coupling strength \(V_{\text{opt}} = 1.4637\kappa\) determined by Eq.~(\ref{2optimal}) in Fig.~\ref{upb_F_V_2}(c)(d).

To demonstrate the mechanism of UPB, Fig.~\ref{level} shows the levels and transition paths. The strong photon antibunching is induced by destructive interference in two-photon excitation processes involving direct and indirect pathways. There are three interference paths occurring in cavity \(a_1\) from $ |00g\rangle$ to $|20g\rangle $: (i) $\left| {00g} \right\rangle \xrightarrow{F} \left| {10g} \right\rangle \xrightarrow{F} \left| {20g} \right\rangle$ through driving field with strength \(F\), (ii) $ |11g\rangle \xrightarrow{\lambda} |20g\rangle $ by intercavity coupling with strength \(\lambda\), and (iii) $ |10e\rangle \xrightarrow{J} |20g\rangle $ via giant atom-cavity coupling with strength \(J\). Similarly, there are two interference paths arising in cavity \(a_2\) from $ |00g\rangle$ to $|02g\rangle $: (i) $\left| {11g} \right\rangle \xrightarrow{\lambda} \left| {02g} \right\rangle $ through intercavity coupling with strength \(\lambda\) and (ii) $\left| {01e} \right\rangle \xrightarrow{V} \left| {02g} \right\rangle $ by giant atom-cavity coupling with strength \(V\) in Fig. \ref{level}. The occurrence of simultaneous UPBs is caused by these destructive interference pathways, which can be derived from Eq. (\ref{C20g}).
% 图10
\begin{figure}[t]
\centering\scalebox{0.41}{\includegraphics{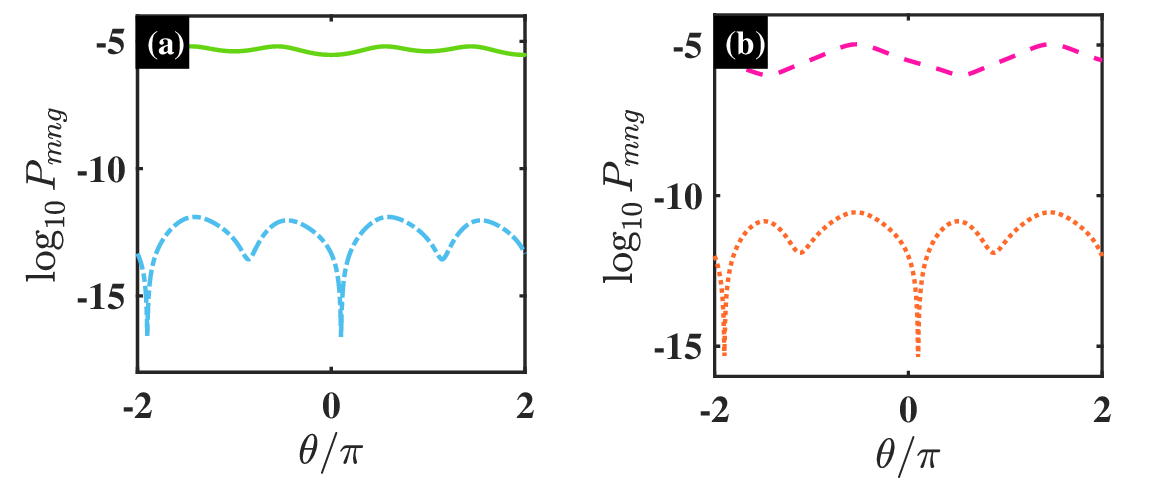}}
\caption{State occupations \(\log_{10}P_{mng}\) (solved by Eq.~(\ref{rou2})) with giant atom and two cavities system as a function of phase $\theta$. (a) Single- and two-photon states probabilities $P_{10g}$ and $P_{20g}$ for cavity $a_1$ versus phase $\theta$. (b) Single- and two-photon states probabilities $P_{01g}$ and $P_{02g}$ for cavity $a_2$ versus phase $\theta$. The parameters chosen are the same as those in Fig.~\ref{one}(a)(c).}
\label{Pab}
\end{figure}
% 图11
\begin{figure}[b]
\centering\scalebox{0.23}{\includegraphics{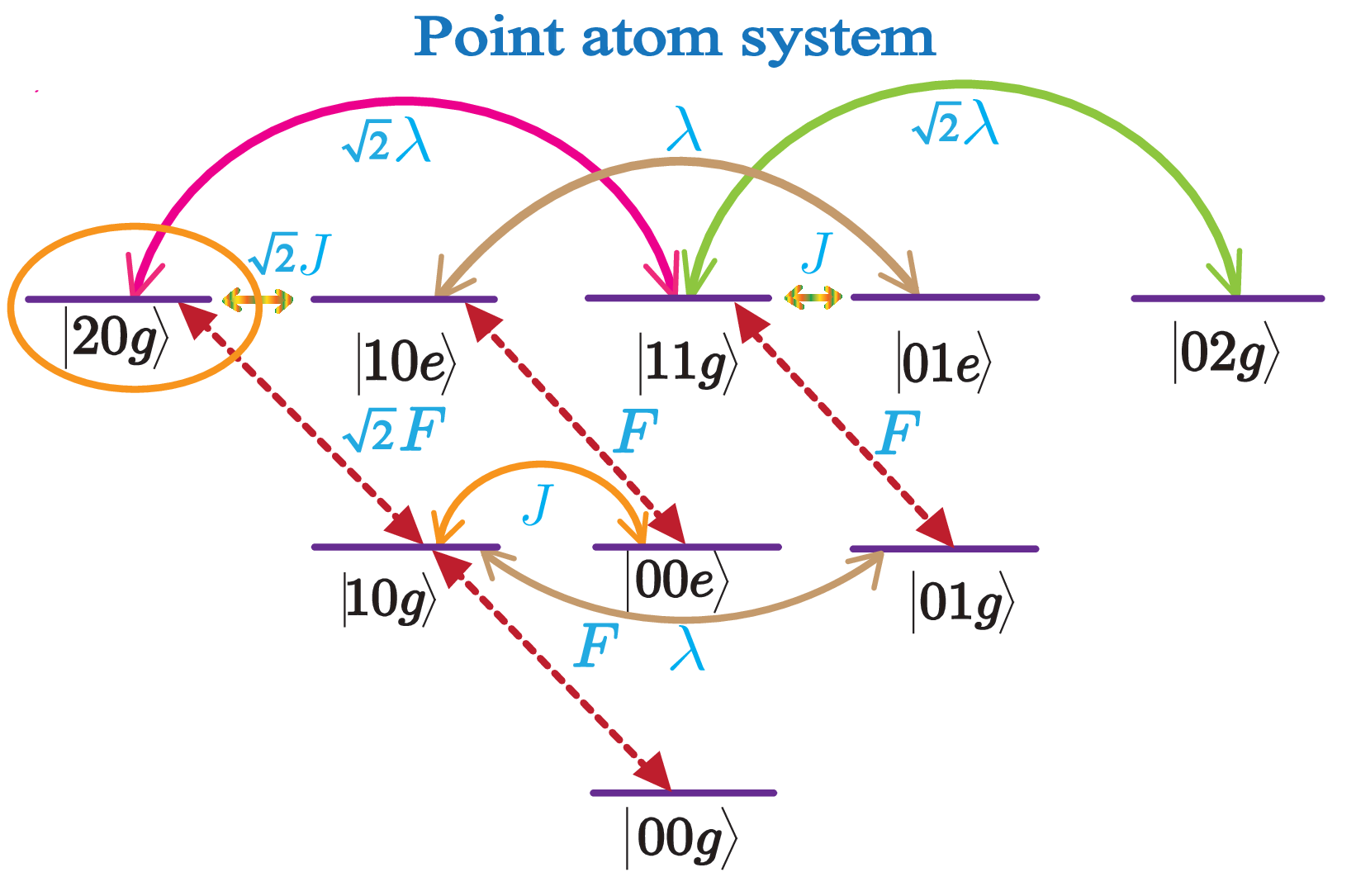}}
\caption{Level diagram of point atom coupled to two cavities system. The absence of simultaneous UPBs is due to the lack of a destructive interference path in cavity $a_2$ caused by the single-point coupling.}
%\vspace{0.3cm}
\label{smallatomlevel}
\end{figure}

Figure~\ref{shuzhijiexi} compares the analytical and numerical results for \(g^{(2)}(0)\) as a function of phase \(\theta\). The red solid lines represent the analytical result~(\ref{2g20}), while the blue circles correspond to the numerical simulation by solving the master equation~(\ref{rou2}). The analytical solution shows consistent agreement with the numerical result in Fig.~\ref{shuzhijiexi}, confirming the validity of our approach. A minor deviation in \(g^{(2)}(0)\) can be observed in Fig.~\ref{shuzhijiexi}, which is attributable to the non-Hermitian Schr\"odinger equation approach truncating at two photons and neglecting multiphoton contributions.

The significant difference in probability between single- and two-photon states is an indicator for obtaining photon blockade \cite{ZYJin0537022023,ZYJin0124592024}. Figure~\ref{Pab} shows the single- and two-photon probability distributions. We observe a suppression of the populations for states $|20g\rangle$ and $|02g\rangle$ at the optimal phase $\theta_{\mathrm{opt}}$ given by Eq.~(\ref{2optimal}). Owing to the existence of multiple destructive interference paths, we find a high single-photon probability together with a strongly suppressed two-photon state in Fig.~\ref{Pab}. At $\theta_{\mathrm{opt}}$ obtained by Eq.~(\ref{2optimal}), the two-photon probabilities \(P_{20g}\) and \(P_{02g}\) are suppressed to \( 10^{-15}\), whereas the single-photon probabilities \(P_{10g}\) and \(P_{01g}\) maintain \(10^{-5}\) in Fig.~\ref{Pab}. The ten-orders-of-magnitude suppression at $\theta_{\mathrm{opt}}$ demonstrates the generation of high-purity single-photon state.
\section{Simultaneous UPBs, simultaneous CPBs, and simultaneous 2PBs for point atom and two cavities system}
\label{si}
In this section, we investigate simultaneous UPBs, simultaneous CPBs, and simultaneous 2PBs in two cavities for the point atom configuration by setting $V=0$ and $\theta=0$ in Eq.~(\ref{H2}). The corresponding master equation is
\begin{align}
 \dot \rho_0 = - i[\hat H_0,\rho_0 ] +  \sum_{j=1}^{2} \kappa {{\cal L}(a_j)}\rho_0 + \kappa {{\cal L}(\sigma)}\rho_0 , \label{rou2point}
\end{align}
where $\hat H_0 = {\Delta_1}{\hat a}_1^\dag {{\hat a}_1} + {\Delta_2}{\hat a}_2^\dag {{\hat a}_2} + {\Delta_0}{\sigma ^ + }\sigma + J({{\hat a}_1}{\sigma ^ + } + {\hat a}_1^\dag \sigma )+ \lambda ({{\hat a}_1}{\hat a}_2^\dag + {{\hat a}_2}{\hat a}_1^\dag )+ F({{\hat a}_1} + {\hat a}_1^\dag )$. Figure \ref{smallatomlevel} shows the level diagram for point atom coupled to two cavities (the atom coupling only to the leftmost cavity). The two-photon state \(|20g\rangle\) in cavity \(a_1\) arises from the same three interference paths as those described for the giant atom in Sec.~\ref{sanB}. However, the transition from $|00g\rangle$ to $|02g\rangle$ in cavity $a_2$ occurs only through $|11g\rangle \xrightarrow{\lambda} |02g\rangle$ in Fig.~\ref{smallatomlevel}. The absence of such destructive interference pathways deprives cavity $a_2$ of the optimal condition to get UPB, which makes the two-photon occupation in state $|02g\rangle$ inevitable. Consequently, the point atom system cannot obtain simultaneous UPBs in two cavities.

In contrast, the multi-point coupling of the giant atom system enables destructive interference between $|01e\rangle \xrightarrow{V} |02g\rangle$ and $|11g\rangle \xrightarrow{\lambda} |02g\rangle$ transition paths in cavity \(a_2\) in Fig.~\ref{level}, which suppresses the two-photon population and leads to UPB. Using the optimal parameters determined by Eq.~(\ref{2optimal}), we realize simultaneous UPBs in two cavities in Fig. \ref{one}.

Moreover, by setting $\lambda=J$ and ${\omega _1} = {\omega _2} = {\omega _0} \equiv \omega' $ in Eqs.~(\ref{H2}) and (\ref{H3}) without the driving terms, the eigenfrequencies of the point atom system in single- and two-excitation subspaces are obtained as
\begin{equation}
\begin{aligned}
        \omega_{1}^{(1)} &= {\omega'}, \quad \omega_{2,3}^{(1)} = {\omega'} \pm \sqrt{2}J, \quad \omega_1^{(2)} = 2 {\omega'}, \\
        \omega_{2,3}^{(2)} &= 2 {\omega'} \pm \sqrt{4 + \sqrt{10}} \, J, \quad \omega_{4,5}^{(2)} = 2 {\omega'} \pm \sqrt{4 - \sqrt{10}} \, J,
\label{pointatomomega}
\end{aligned}
\end{equation}
from which the optimal detunings for single- and two-excitation subspaces are respectively written as
\begin{equation}
\begin{aligned}
\Delta_{\text{opt}}  = \, 0\quad\text{or}\quad\Delta_{\text{opt}}  = \pm \sqrt{2} J,
\label{2Delta2_point_1}
\end{aligned}
\end{equation}
and
\begin{equation}
\begin{aligned}
\Delta_{\text{opt}}  &= \, 0,\,\,\,\,\,\,\,\,\,\,\,\,\,\,\,\Delta_{\text{opt}}  = \pm \frac{1}{2}\sqrt{4 + \sqrt{10}} \, J,\\
\Delta_{\text{opt}}&= \pm \frac{1}{2}\sqrt{4 - \sqrt{10}} \, J.
\label{2Delta2_point}
\end{aligned}
\end{equation}
% 图12
\begin{figure}[t]
\centering\scalebox{0.39}{\includegraphics{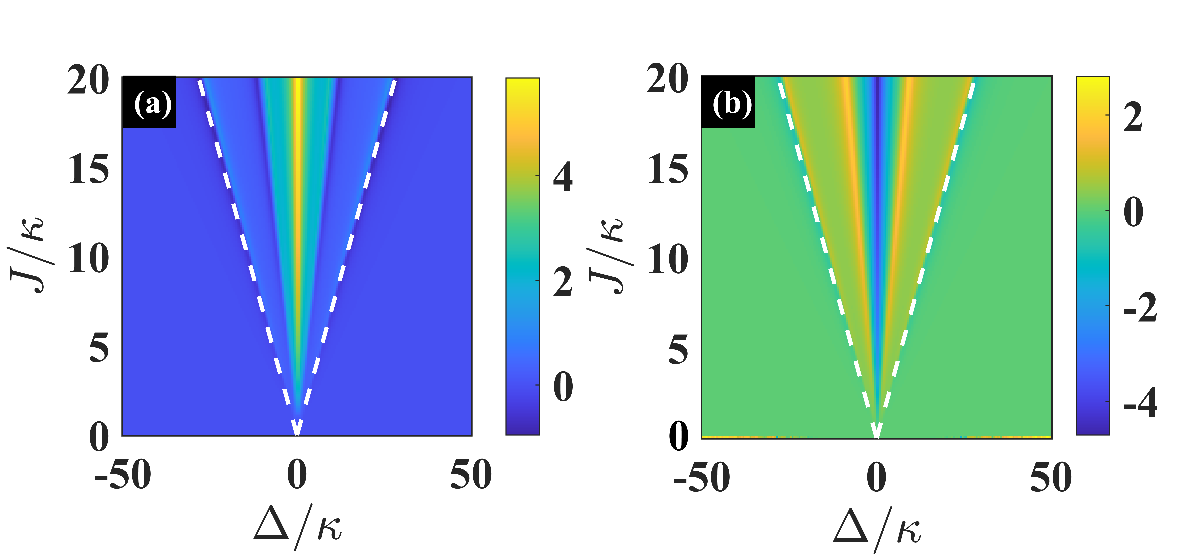}}
\caption{\(\log_{10}g^{(2)}_{o}(0)\) (solved by Eq.~(\ref{rou2point})) with point atom and two cavities system as a function of detuning \( \Delta \) and coupling strength \( J \) with \( a_1 \) in (a) and \( a_2 \) in (b). CPBs occur simultaneously in two cavities at the same location at \(\Delta_{\text{opt}} = \pm \sqrt{2} J\) (corresponding to four white dashed lines) for (a)(b). The parameters chosen are $J=20\kappa$ and \(V=0\), while other parameters chosen are the same as those in Fig.~\ref{CPB}.}
\label{pointatomCPB}
\end{figure}
%图13
\begin{figure}[t]
\centering\scalebox{0.37}{\includegraphics{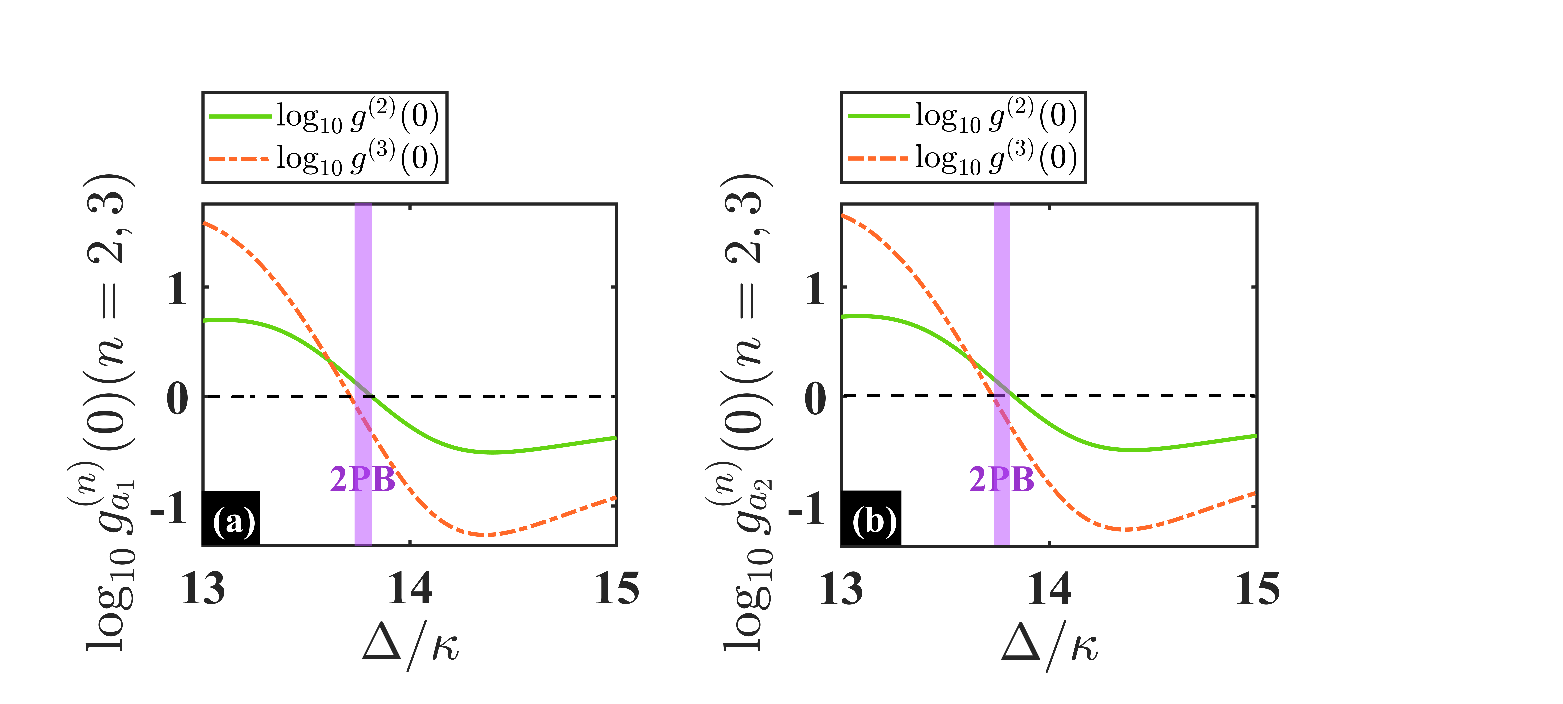}}
\caption{\( \log_{10}g^{(2)}_{o}(0) \) (green solid line) and \( \log_{10}g^{(3)}_{o}(0) \) (orange dot-dashed line) (solved by Eq.~(\ref{rou2point})) versus detuning $\Delta$ with $a_1$ in (a) and $a_2$ in (b) for point atom and two cavities system. The black dashed lines indicate the reference value 1 for $g^{(2)}_{o}(0)$ and $g^{(3)}_{o}(0)$. The purple shaded regimes satisfy $g^{(2)}(0) > 1$ and $g^{(3)}(0) < 1$ as defined in Eq.~(\ref{2g20g30}) (corresponding to simultaneous 2PBs in two cavities). The parameters chosen are \(J=10\kappa\), \(V=0\), $\theta=0$, and $F=0.1\kappa$.}
\label{pointatom2PB}
\end{figure}

When resonance occurs \(\omega_l = \omega_{2,3}^{(1)}\) (i.e., \(\Delta_{\text{opt}} = \pm \sqrt{2} J\) in Eq.~(\ref{2Delta2_point_1})), simultaneous CPBs can be observed in point atom and two cavities system by four white dashed lines in Fig. \ref{pointatomCPB}.

In addition, we study 2PB in point atom and two cavities system. In single- and two-excitation subspaces, we can obtain the corresponding eigenfrequencies in Eq.~(\ref{pointatomomega}).

2PB denotes the phenomenon where two-photon absorption is permitted but three-photon absorption is suppressed \cite{LJFeng0435092021,QBin0438582018,WWDeng0438312015,AMiranowicz0238092013,AKowalewskaKudlaszyk0538572019,QCWu212022}. In this quantum optical process, nonlinear interactions within the system keep the two-photon transition channels open while inhibiting three-photon transitions, which has been experimentally observed \cite{CHamsen1336042017}. Here we show the requirements of 2PB on correlation functions
\begin{equation}
\begin{aligned}
g_{{o}}^{(2)}(0) = \frac{\langle \hat{o}^\dagger \hat{o}^\dagger \hat{o} \hat{o} \rangle}{\langle \hat{o}^\dagger \hat{o} \rangle^2} > 1,\, g_{{o}}^{(3)}(0) = \frac{\langle \hat{o}^\dagger \hat{o}^\dagger \hat{o}^\dagger \hat{o} \hat{o} \hat{o} \rangle}{\langle \hat{o}^\dagger \hat{o} \rangle^3} < 1,
\label{2g20g30}
\end{aligned}
\end{equation}
where $\hat{o} = \hat{a}_1$ and $\hat{a}_2$. Near the two-photon eigenfrequencies, it is possible to observe 2PB (i.e., \(g^{(2)}(0)> 1\) and \(g^{(3)}(0)<1\) in Eq.~(\ref{2g20g30})). The point atom and two cavities system exhibits simultaneous 2PBs in two cavities by the purple shaded regions near two-photon resonance condition $\Delta_{\text{opt}} = \sqrt{4 + \sqrt{10}} \, J / 2 \approx 1.3381J \approx 13.381 \kappa$ originated from Eq.~(\ref{2Delta2_point}) in Fig.~\ref{pointatom2PB}(a)(b).
\section{The effect of giant atom on simultaneous photon blockades}
\label{wu}
\subsection{CPBs occur simultaneously in three cavities}
\label{wuA}
In this section, we investigate influences of giant atom effects on simultaneous photon blockades with giant atom coupling to three cavities in Eq.~(\ref{H3}). Assuming $\theta=0$, $ \lambda  = J \equiv V$, and $\omega_1 = \omega_2 = \omega_3 = \omega_0 \equiv \omega'$ (or $\Delta_1 = \Delta_2 = \Delta_3  = \Delta_0 \equiv \Delta$), the matrix form of Hamiltonian (\ref{H3}) without the driving terms in single-excitation subspace can be written as
% 图14
\begin{figure}[b]
\centering{\includegraphics[width=8cm,  height=5cm,  clip]{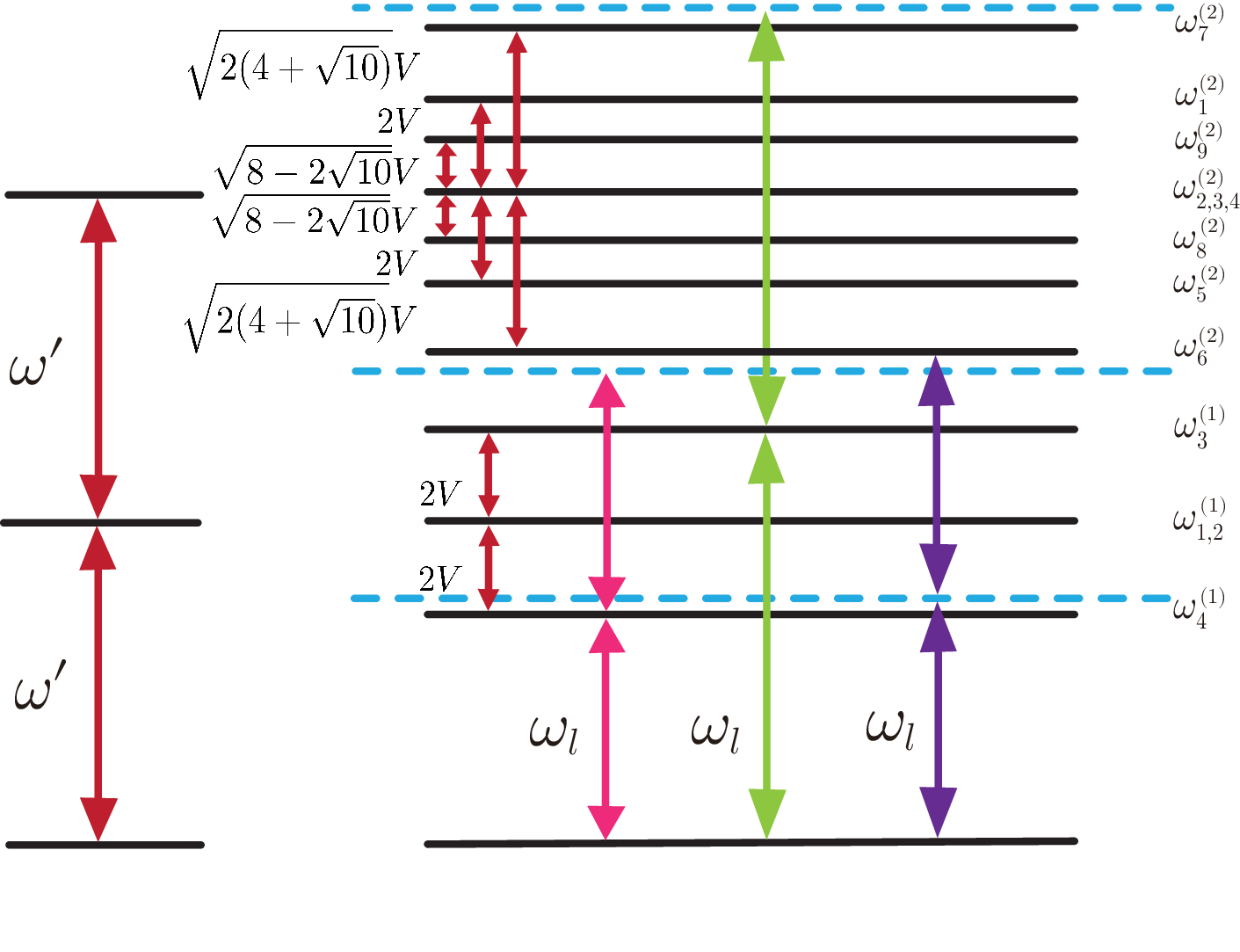}}
\caption{Level diagram of giant atom and three cavities system illustrates CPB mechanism. The horizontal lines correspond to the eigenfrequencies. The frequency $\omega'$ denotes the common bare frequency of the cavities and giant atom. The pink and green solid arrows indicate single-photon resonance at \(\Delta_{\text{opt}} = \pm 2V\) in Eq.~(\ref{3Delta1}), while the purple solid arrow represents two-photon resonance at \(\Delta_{\text{opt}} = \sqrt{2(4+\sqrt{10})}V/2\approx1.8924V\) obtained by Eq.~(\ref{3Delta3}).}
\label{CPBlevel3}
\end{figure}
\begin{equation}
\hat H'_3 = \left( {\begin{array}{*{20}{c}}
{{\omega'}}&V &0&V\\
V &{\omega'}&V &0\\
0&V &{\omega'}&{V}\\
V&0&{V}&{\omega'}
\end{array}} \right),
   \label{H3CPB1}
\end{equation}
with corresponding eigenfrequencies $\omega_{1,2}^{(1)} = \omega'$, $\omega_{3,4}^{(1)} = \omega' \pm 2V$ and optimal detunings
\begin{align}
\Delta_{\text{opt}} = 0 \quad\text{or}\quad \Delta_{\text{opt}} = \pm 2V.
 \label{3Delta1}
\end{align}
Within the two-excitation subspace, the matrix form of the Hamiltonian (\ref{H3}) without the driving terms reads
\begin{widetext}
\begin{align}
\hat H''_3 = \left( {\begin{array}{*{20}{c}}
{2{\omega'}}&0&0&{\sqrt 2 V }&0&0&{\sqrt 2 V}&0&0\\
0&{2{\omega'}}&0&{\sqrt 2 V }&0&{\sqrt 2 V }&0&0&0\\
0&0&{2{\omega'}}&0&0&{\sqrt 2 V }&0&0&{\sqrt 2 V}\\
{\sqrt 2 V }&{\sqrt 2 V }&0&{2{\omega'} }&V &0&0&V&0\\
0&0&0&V &{2{\omega '} }&V &{V}&0&V\\
0&{\sqrt 2 V }&{\sqrt 2 V }&0&V &{2{\omega '} }&0&{V}&0\\
{\sqrt 2 V}&0&0&0&{V}&0&{2{\omega '} }&V &0\\
0&0&0&V&0&{V}&V &{2{\omega'}}&V \\
0&0&{\sqrt 2 V}&0&V&0&0&V &{2{\omega'} }
\end{array}} \right),
 \label{3Delta2}
\end{align}
\end{widetext}
whose  eigenfrequencies can be derived as
% 图15
\begin{figure}[t]
\centering
{\includegraphics[width=8.5cm,  height=3cm,  clip]{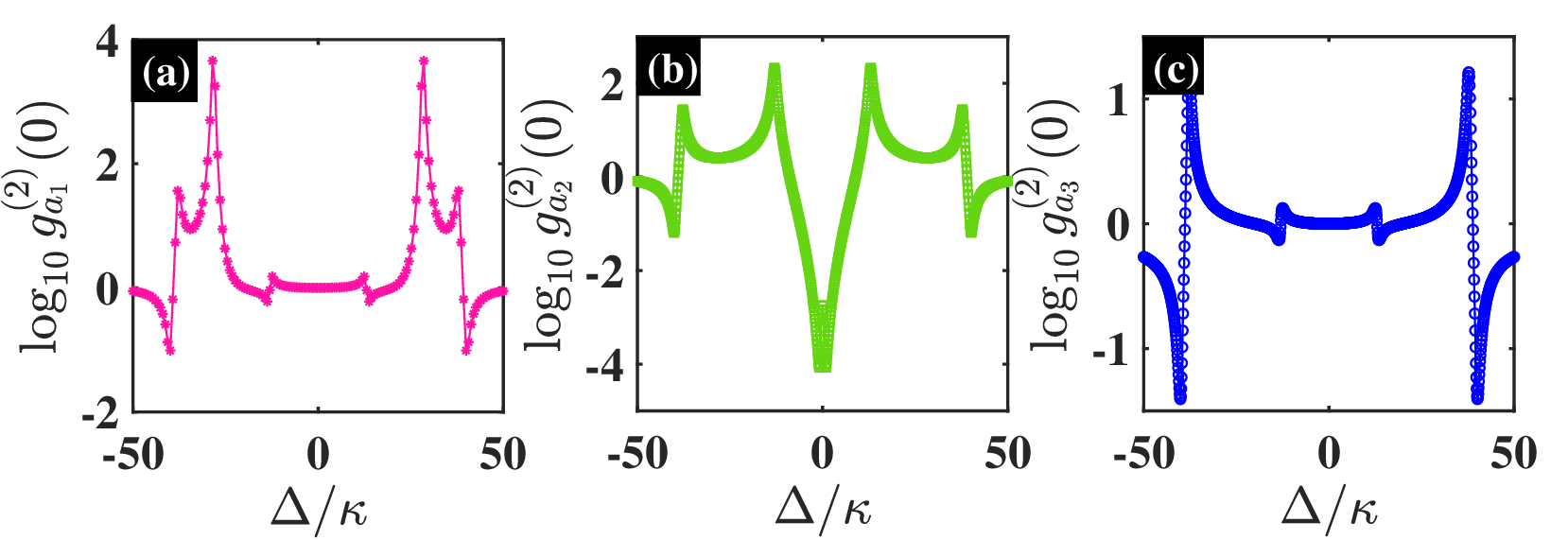}}
\caption{\( \log_{10}g^{(2)}_{o}(0) \) (solved by Eq.~(\ref{rou3})) with giant atom and three cavities system versus detuning $\Delta$ with $a_1$ in (a), $a_2$ in (b), and $a_3$ in (c). CPBs occur simultaneously in three cavities at the same location at \( \Delta_{\text{opt}} = \pm 2V \) for (a)-(c) given by Eq.~(\ref{3Delta1}). The parameter chosen is $V=20\kappa$, while other parameters chosen are the same as those in Fig. \ref{CPB}}
\label{CPB3}
\end{figure}
\begin{equation}
\begin{aligned}
\omega _{1,5}^{(2)} &= 2{\omega'} \pm 2V,\\
\omega _{2,3,4}^{(2)} &= 2{\omega'},\\
\omega _{6,7}^{(2)} &= 2{\omega'} \pm \sqrt {2(4 + \sqrt {10} )} V,\\
\omega _{8,9}^{(2)} &= 2{\omega'} \pm \sqrt {8 - 2\sqrt {10} } V.
 \label{3omega2}
\end{aligned}
\end{equation}
From Eq.~(\ref{3omega2}), we can obtain the optimal detunings
\begin{equation}
\begin{aligned}
\Delta_{\text{opt}}  =&\pm V,\,\,\,\,\,\,\,\,\,\,\Delta_{\text{opt}}  = 0,\,\,\,\,\,\,\,\,\,\,\,\Delta_{\text{opt}}  = \pm \frac{1}{2}\sqrt {2(4 + \sqrt {10} )} V,\\
\Delta_{\text{opt}}=& \pm \frac{1}{2}\sqrt {8 - 2\sqrt {10} } V.
\label{3Delta3}
\end{aligned}
\end{equation}

Figure \ref{CPBlevel3} illustrates the physical mechanism of CPB through the transition pathways. When an incident laser at frequency $\omega_l$ drives the system, a photon resonant with the cavity frequencies $\omega_{3}^{(1)}$ and $\omega_{4}^{(1)}$ excite three cavities from the vacuum state $|000\rangle$ to the first excited states $|100\rangle$, $|010\rangle$, and $|001\rangle$. However, once every one of three cavities holds a single photon simultaneously, the absorption of a second photon (transitions $\omega_{3}^{(1)} \rightarrow \omega_{7}^{(2)}$ and $\omega_{4}^{(1)} \rightarrow \omega_{6}^{(2)}$) is suppressed as a result of a large detuning \( (4 - \sqrt{2(4 + \sqrt{10})} ) V \approx 0.2152V\). As indicated by the solid pink and green arrows in Fig. \ref{CPBlevel3}, the optimal condition in Eq. (\ref{3Delta1}) corresponds to CPB, where the system cannot be excited to the second energy level.

Figure \ref{CPB3} presents \( g^{(2)}(0) \) as a function of detuning \(\Delta\) in three cavities system. In Fig.~\ref{CPB3}(a)-(c), \(g^{(2)}(0)\) of three cavities reaches minimum simultaneously at \( \Delta_{\text{opt}} = \pm 2V \) from Eq.~(\ref{3Delta1}), which results from energy-level anharmonicity and indicates simultaneous presence of CPBs in three cavities.
\subsection{UPBs occur simultaneously in three cavities}
\label{wuB}
In this section, we analyze simultaneous UPBs in Eq.~(\ref{H3}) by setting $\Delta_1 = \Delta_2  \equiv \Delta \neq \Delta_3 \neq \Delta_0$. The effective Hamiltonian reads
\begin{equation}
\hat{H}'_{\text{eff}} = \hat H_3 - \frac{i}{2}\kappa \hat a_1^\dagger \hat a_1 - \frac{i}{2}\kappa \hat a_2^\dagger \hat a_2 - \frac{i}{2}\kappa \hat a_3^\dagger \hat a_3 - \frac{i}{2}\kappa \sigma^+ \sigma.
\label{H3eff}
\end{equation}
The state in Eq.~(\ref{2pusai}) is
% 图16
\begin{figure}[t]
\centering\scalebox{0.315}{\includegraphics{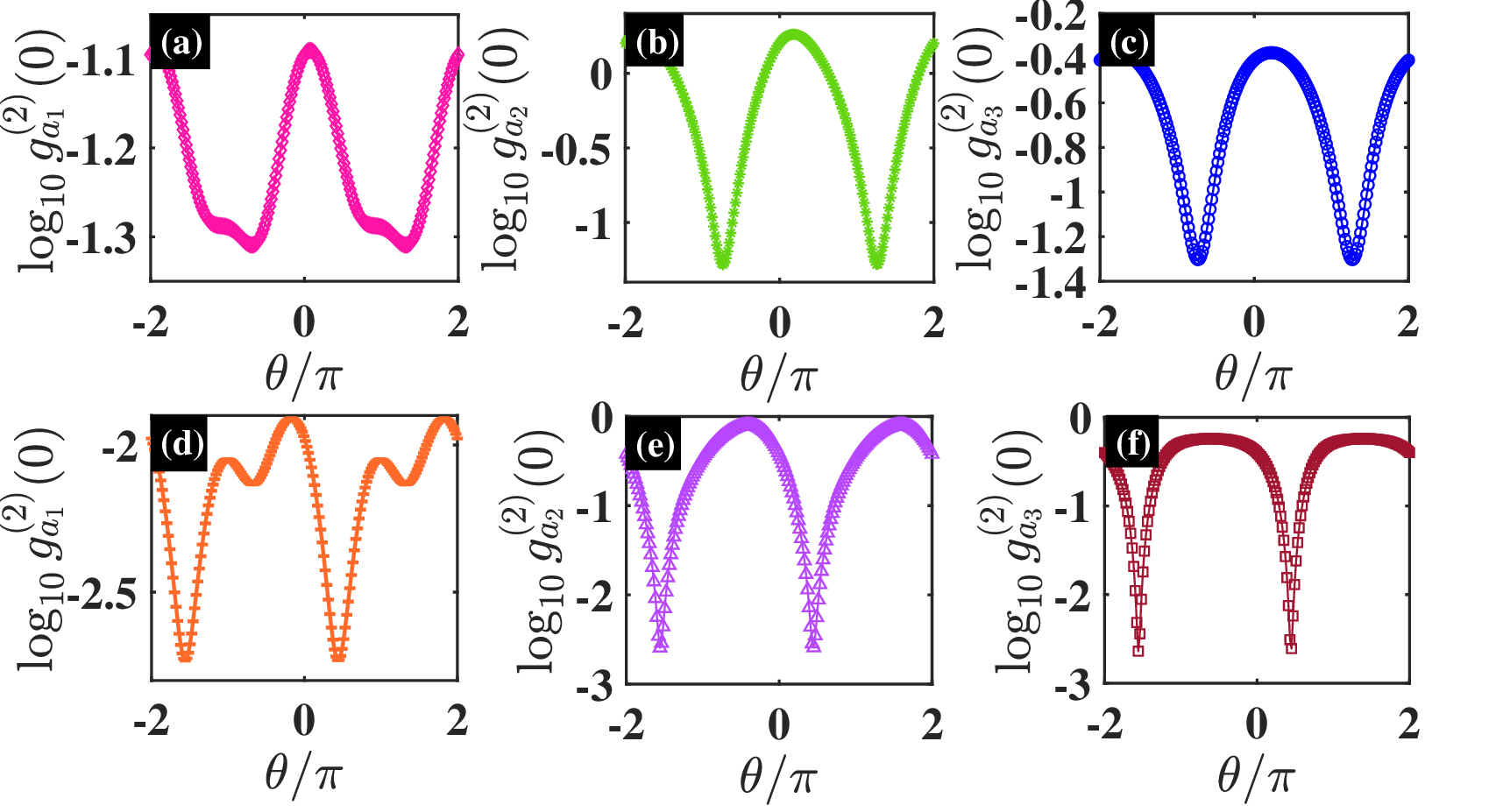}}
\caption{\( \log_{10}g^{(2)}_{o}(0) \) (solved by Eq.~(\ref{rou3})) with giant atom and three cavities system versus phase $\theta$ with $a_1$ in (a)(d), $a_2$ in (b)(e), and $a_3$ in (c)(f). (a)-(c) shows results for the first parameters set $\lambda = 0.6872\kappa$, $J_{\text{opt}} = 0.7996\kappa$, ${\Delta_3}_{\text{opt}}=-0.1195\kappa$, $V_{\text{opt}} = 0.0627\kappa$, \(F=0.05\kappa\), ${\Delta_0}_{\text{opt}} = 0.0999\kappa$, and $\Delta_{\text{opt}} = -0.403\kappa$. (d)-(f) displays results for the second set $\lambda = 0.6558\kappa$, $J_{\text{opt}} = 0.8678\kappa$, $V_{\text{opt}} = 0.0507\kappa$, ${\Delta_3}_{\text{opt}} = 0.2105\kappa$, \(F=0.01\kappa\), ${\Delta_0}_{\text{opt}} = -0.2306\kappa$, and $\Delta_{\text{opt}} = 0.9242\kappa$ determined by Eq.~(\ref{3optimal}). UPBs occur simultaneously in three cavities at the same location at \(\theta_{\text{opt}} = 1.2732\pi\) (with period 2$\pi$) for (a)-(c), also at \(\theta_{\text{opt}} =0.4456\pi\) (with period 2$\pi$) for (d)-(f).}
\label{threetheta}
\end{figure}
\begin{align}
\left|  \Psi \right\rangle = & {C_{000g}}\left| {000g} \right\rangle + {C_{100g}}\left| {100g} \right\rangle + {C_{010g}}\left| {010g} \right\rangle \nonumber\\
& + {C_{001g}}\left| {001g} \right\rangle + {C_{000e}}\left| {000e} \right\rangle + {C_{200g}}\left| {200g} \right\rangle \nonumber\\
& + {C_{020g}}\left| {020g} \right\rangle + {C_{002g}}\left| {002g} \right\rangle + {C_{110g}}\left| {110g} \right\rangle \nonumber\\
& + {C_{101g}}\left| {101g} \right\rangle + {C_{011g}}\left| {011g} \right\rangle + {C_{100e}}\left| {100e} \right\rangle \nonumber\\
& + {C_{010e}}\left| {010e} \right\rangle + {C_{001e}}\left| {001e} \right\rangle.
\label{3pusai}
\end{align}
Under the weak driving condition, we derive the amplitudes \(C_{200g}\), \(C_{020g}\), and \(C_{002g}\) in Appendix \ref{B}. From Eqs. (\ref{go20}) and (\ref{3pusai}), we obtain
\begin{small}
\begin{equation}
\begin{aligned}
g_{{a}_1}^{(2)}(0)  \simeq \frac{2 |C_{200g}|^2}{|C_{100g}|^4},
g_{{a}_2}^{(2)}(0)  \simeq \frac{2 |C_{020g}|^2}{|C_{010g}|^4},
g_{{a}_3}^{(2)}(0)  \simeq \frac{2 |C_{002g}|^2}{|C_{001g}|^4}.
\label{3g20}
\end{aligned}
\end{equation}
\end{small}Simultaneous UPBs in three cavities require optimal condition
% 图17
\begin{figure}[t]
\centering\scalebox{0.29}{\includegraphics{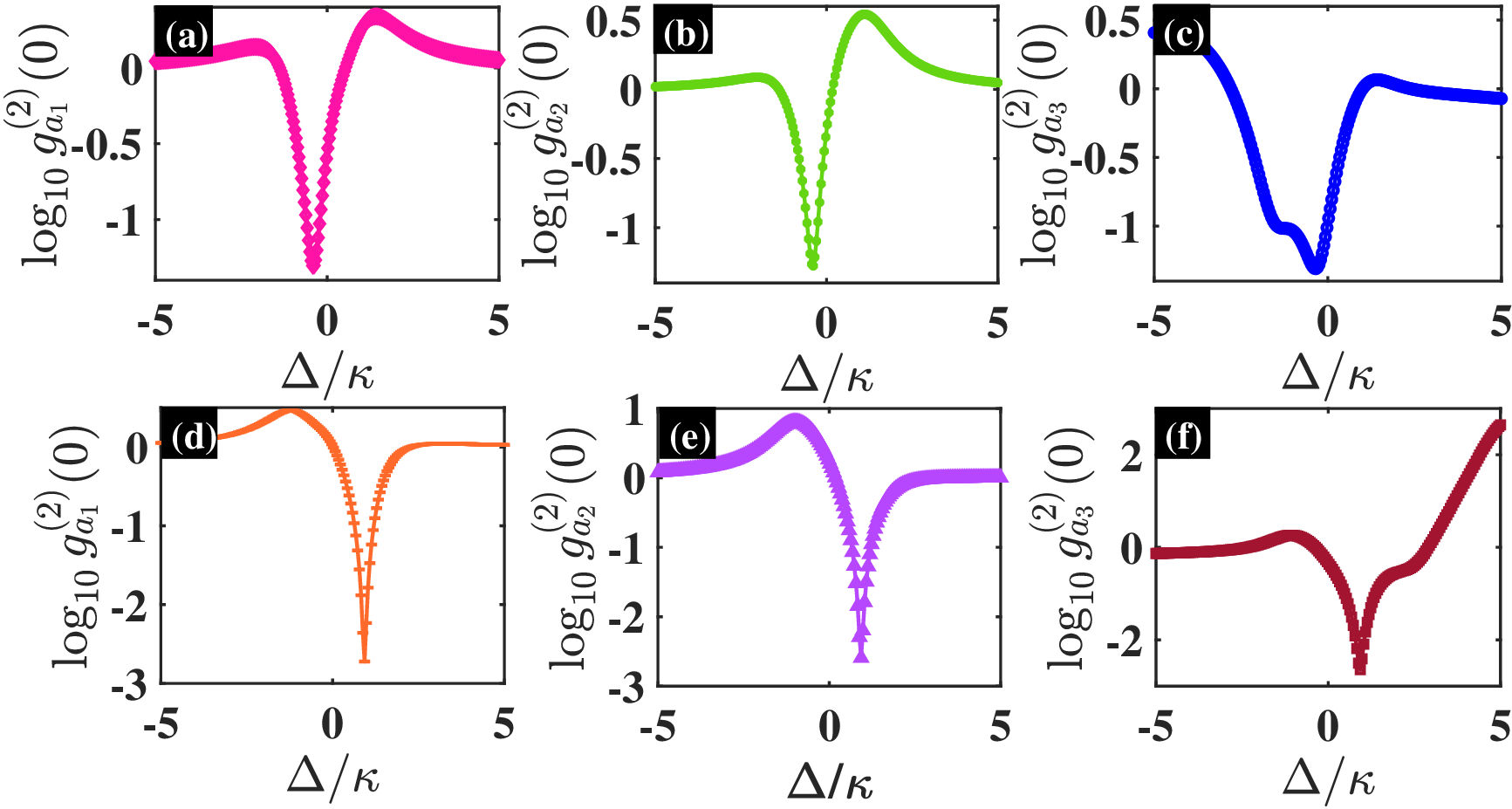}}
\caption{\( \log_{10}g^{(2)}_{o}(0) \) (solved by Eq.~(\ref{rou3})) with giant atom and three cavities system versus detuning $\Delta$ with $a_1$ in (a)(d), $a_2$ in (b)(e), and $a_3$ in (c)(f), where (a)-(c) corresponds to the optimal phase $\theta_{\text{opt}} = 1.2732\pi$, while (d)-(f) corresponds to $\theta_{\text{opt}} = 0.4456\pi$ obtained by Eq.~(\ref{3optimal}). UPBs occur simultaneously in three cavities at the same location at \(\Delta_{\text{opt}} = -0.403\kappa\) for (a)-(c), also at \(\Delta_{\text{opt}} = 0.9242\kappa\) for (d)-(f). The other parameters chosen are the same as those in Fig.~\ref{threetheta}.}
\label{threeDelta}
\end{figure}
\begin{equation}
\begin{aligned}
{\mathop{\rm Re}\nolimits} (A') &= 0, \,\,\,\,\, {\mathop{\rm Im}\nolimits} (A') = 0,\,\,\,\,\, {\mathop{\rm Re}\nolimits} (C') = 0, \\
{\mathop{\rm Im}\nolimits} (C') &= 0, \,\,\,\,\, {\mathop{\rm Re}\nolimits} (D') = 0,\,\,\,\,\, {\mathop{\rm Im}\nolimits} (D') = 0,
\label{3optimal}
\end{aligned}
\end{equation}
where \(A'\), \(C'\), and \(D'\) are given in Appendix \ref{B}. Utilizing Eq.~(\ref{3optimal}), we plot \( g^{(2)}(0) \) as functions of phase $\theta$ and detuning $\Delta$ in Figs.~\ref{threetheta} and \ref{threeDelta}. Figure \ref{threetheta}(a)-(c) demonstrates simultaneous minima in \( g^{(2)}_{o}(0) \) at \( \theta_{\text{opt}} = 1.2732\pi \) (with period 2$\pi$) obtained by Eq.~(\ref{3optimal}), which indicates the occurrence of simultaneous UPBs in three cavities. The system also supports simultaneous UPBs in three cavities at $\theta_{\text{opt}} = 0.4456\pi$ (with period 2$\pi$) given by Eq.~(\ref{3optimal}) in Fig.~\ref{threetheta}(d)-(f). Figure~\ref{threeDelta}(a)-(c) demonstrates that simultaneous UPBs arise in three cavities at \(\Delta_{\text{opt}} = -0.403\kappa\) determined by Eq.~(\ref{3optimal}). Similarly, another set of parameters obtained by Eq.~(\ref{3optimal}) yields simultaneous UPBs in three cavities at \(\Delta_{\text{opt}} = 0.9242\kappa\) in Fig.~\ref{threeDelta}(d)-(f).

Moreover, we compare UPB coupled to either point or giant atom. Due to the single-point coupling between point atom and cavity (the atom coupling only to the leftmost cavity), the two-photon state \(|002g\rangle\) can only be excited via a single transition pathway, i.e., \(|011g\rangle \xrightarrow{\lambda} |002g\rangle\) governed by Eq.~(\ref{C200g}). The absence of destructive interference pathways leads to the occupation of two-photon state in cavity \(a_3\), which prevents simultaneous formation of UPBs in three cavities. However, in the giant atom system with multipoint coupling, the two-photon state \(|002g\rangle\) exhibits two different destructive interference pathways: \(|011g\rangle \xrightarrow{\lambda} |002g\rangle\) and \(|001e\rangle \xrightarrow{V} |002g\rangle\). These pathways collectively facilitate simultaneous occurrence of UPBs in three cavities.
\section{conclusions}
\label{liu}
In summary, we have investigated simultaneous photon blockades in giant atom coupled to two cavities and three cavities systems with the first cavity mediated by driving field. The point atom system in our configuration (with a point atom only coupled to leftmost cavity) cannot produce simultaneous UPBs in multiple cavities due to the absence of destructive interference pathways. However, the giant atom system overcomes this constraint through multipoint coupling and leads to simultaneous UPBs in multiple cavities under the optimal conditions. The introduced phase $\theta$ and coupling strength $V$ provide additional degrees of freedom that facilitate the occurrence of simultaneous UPBs. Moreover, simultaneous CPBs were obtained at the single-photon resonance in multiple cavities owing to energy-level anharmonicity. Simultaneous 2PBs have also been investigated in point atom multiple cavities system. This work demonstrates giant atom mediated two cavities and three cavities systems that presents a new perspective for engineering simultaneous photon blockades through other systems, e.g., (1) Rabi model $\sum\nolimits_j {{G_j}{\sigma _x}} (\hat a_j^\dag  + {{\hat a}_j})$ \cite{HZShen0121072016,YHChen52024,YHChen0336032024}, (2) first-order coupling ${{\hat a}^\dag }\hat a(\hat c + {{\hat c}^\dag })$ \cite{WZZhang0638532016,WZhang0337012024,JFTriana1836022016,TJ2024}, (3) quadratic optomechanical couplings ${\hat a^\dag }\hat a{(\hat b + {\hat b^\dag })^2}$ \cite{JCSankey7072010,MBhattacharya0338192008}, (4) Kerr and cross-Kerr nonlinear medium ${\hat a^\dag }{\hat a^\dag }\hat a\hat a$ \cite{ SFerretti0333032012,HZShen0638082015,dl1,dl2,dl3}.
\section*{ACKNOWLEDGMENTS}

We thank the referees for their constructive comments that helped improve our work. H.Z.S. and C.C. would like to thank Dr.~Hai-Jun Xing and Dr.~Gangcheng Wang for valuable and constructive discussions. This work was supported by Science and Technology Development Plan Project of Jilin Province (Grant No. 20250102007JC), National Natural Science Foundation of China under Grant No. 12274064 and No. 12374333, and Shandong Provincial Natural Science Foundation under Grant No. ZR2021MA036.
\section*{DATA AVAILABILITY}

The data that support the findings of this article are not publicly available. The data are available from the authors upon reasonable request.
\appendix
\begin{widetext}
\section{Derivation of \(C_{20g}\) and \(C_{02g}\) in Eq. (\ref{2g20})}
\label{A}
Through Schr\"{o}dinger equation $i{\partial _t}\left| \psi  \right\rangle  = {\hat H_{\text{eff}}}\left| \psi  \right\rangle  $ with Eqs. (\ref{H2eff}) and (\ref{2pusai}), we obtain
\begin{equation}
\begin{aligned}
i\dot{C}_{10g} &= F{C_{00g}} + \Delta'_1{C_{10g}} + \lambda {C_{01g}} + J{C_{00e}} = 0, \\
i\dot{C}_{01g} &= \lambda {C_{10g}} + \Delta'_2{C_{01g}} + Ve^{i\theta}{C_{00e}} = 0, \\
i\dot{C}_{00e} &= J{C_{10g}} + \Delta'_0{C_{00e}} + Ve^{-i\theta}{C_{01g}} = 0,\\
i\dot{C}_{20g} &= \sqrt{2} F{C_{10g}} + 2\Delta'_1{C_{20g}} + \sqrt{2} \lambda {C_{11g}} + \sqrt{2} J{C_{10e}} = 0, \\
i\dot{C}_{02g} &= 2\Delta'_2{C_{02g}} + \sqrt{2} \lambda {C_{11g}} + \sqrt{2} Ve^{i\theta}{C_{01e}} = 0, \\
i\dot{C}_{11g} &= F{C_{01g}} + \sqrt{2} \lambda {C_{20g}} + \sqrt{2} \lambda {C_{02g}} + \Delta'_1{C_{11g}} + \Delta'_2{C_{11g}} + Ve^{i\theta}{C_{10e}} + J{C_{01e}} = 0, \\
i\dot{C}_{01e} &= \sqrt{2} Ve^{-i\theta}{C_{02g}} + J{C_{11g}} + \lambda {C_{10e}} + \Delta'_2{C_{01e}} + \Delta'_0{C_{01e}} = 0, \\
i\dot{C}_{10e} &= F{C_{00e}} + \sqrt{2} J{C_{20g}} + Ve^{-i\theta}{C_{11g}} + \Delta'_1{C_{10e}} + \Delta'_0{C_{10e}} + \lambda {C_{01e}} = 0.
\label{C20g}
\end{aligned}
\end{equation}
Assuming $\Delta'_1 = \Delta'_2 = \Delta'_0 \equiv Q$ and under the weak driving condition, the probability amplitudes are derived as
\begin{equation}
\begin{aligned}
C_{10g} &= \frac{2 e^{i \theta} F (4 V^2 - (2Q)^2) }{8 J V \lambda + 8 e^{2 i \theta} J V \lambda -e^{i \theta} (2Q) (4 J^2 + 4 V^2 - 4 \Delta^2 + 4 i \Delta \kappa + \kappa^2 + 4 \lambda^2) },  \\
C_{01g} &= \frac{-4 e^{i \theta} F (2 e^{i \theta} J V - 2 \Delta \lambda + i \kappa \lambda) }{8 J V \lambda + 8 e^{2 i \theta} J V \lambda -e^{i \theta} (2Q) (4 J^2 + 4 V^2 - 4 \Delta^2 + 4 i \Delta \kappa + \kappa^2 + 4 \lambda^2) },  \\
C_{20g} &= \frac{A}{B}, \,\,\,\,\,\,\,\,\,\,
C_{02g} = \frac{C}{B},
\label{C20021001}
\end{aligned}
\end{equation}
where \(A\), \(B\), and \(C\) are given by
\begin{align}
A =& 2 \sqrt{2} F^2 \{-e^{i \theta} J V (8 J^2 - 3 (2Q)^2) (2Q) \lambda + 3 e^{3 i \theta} J V (2Q)^3 \lambda + 8 J^2 V^2 \lambda^2 + e^{2 i \theta} [8 V^6 + J^2 (8 V^4 - 4 V^2 (2Q)^2 - (2Q)^4) \nonumber\\
&+ 2 J^4 (2Q)^2 - (2Q)^6 + (2Q)^4 \lambda^2 + V^2 (2Q)^2 (7 (2Q)^2 - 8 \lambda^2) + 2 V^4 (-7 (2Q)^2 - 4 \lambda^2)]\},\nonumber\\
B =& (8 J V \lambda + 8 e^{2 i \theta} J V \lambda - e^{i \theta} (2Q) (4 J^2 + 4 V^2 - 4 \Delta^2 + 4 i \Delta \kappa + \kappa^2 + 4 \lambda^2)) [J V \lambda (4 J^2 + 4 V^2 - 5 (2Q)^2 - 4 \lambda^2) + e^{2 i \theta} \nonumber\\
&J V \lambda (4 J^2 + 4 V^2 - 5 (2Q)^2 - 4 \lambda^2) - e^{i \theta} (2Q) (2 J^4 + 2 V^4 + J^2 (4 V^2 - 3 (2Q)^2) - 3 V^2 (2Q)^2 + (2 \Delta - i \kappa)^4 - 5 (2Q)^2 \nonumber\\
&\lambda^2 + 4 \lambda^4)], \nonumber\\
C =& 8 \sqrt{2} e^{i \theta} F^2 \{3 J V (2Q) \lambda^3 + 2 e^{3 i \theta} J^2 V^2 (J^2 + V^2 - 4 \Delta^2 + 4 i \Delta \kappa + \kappa^2 - \lambda^2) - e^{2 i \theta} J V (2Q) \lambda (2 J^2 + 4 V^2 - 4 (2Q)^2+ \nonumber\\
& \lambda^2) + e^{i \theta} \lambda^2 [2 V^4 - V^2 (2Q)^2 - (2Q)^2 (J^2 + (2Q)^2 - \lambda^2)]\},
\label{ABC}
\end{align}
with $Q = \Delta  - \frac{i}{2}\kappa$.
\section{Derivation of \(C_{200g}\), \(C_{020g}\), and \(C_{002g}\) in Eq. (\ref{3g20})}
\label{B}
Substituting Eqs. (\ref{H3eff}) and (\ref{3pusai}) into the Schr\"{o}dinger equation $i{\partial _t}\left| \Psi  \right\rangle  = {\hat H'_{\text{eff}}}\left| \Psi  \right\rangle  $ gets
\begin{equation}
\begin{aligned}
i\dot{C}_{100g} &=\Delta_{11} C_{100g} + \lambda C_{010g} + J C_{000e} + F C_{000g} = 0, \\
i\dot{C}_{010g} &=\lambda C_{100g} + \Delta_{21} C_{010g} + \lambda C_{001g} = 0,\\
i\dot{C}_{001g} &=\lambda C_{010g} + \Delta_{31} C_{001g} + V e^{i\theta} C_{000e} = 0,\\
i\dot{C}_{000e} &=J C_{100g} + V e^{-i\theta} C_{001g} + \Delta_{01} C_{000e} = 0,\\
i\dot{C}_{200g} &=2\Delta_{11} C_{200g} + \sqrt{2}\lambda C_{110g} + \sqrt{2}J C_{100e} + \sqrt{2}F C_{100g} = 0, \\
i\dot{C}_{020g} &=2\Delta_{21} C_{020g} + \sqrt{2}\lambda C_{110g} + \sqrt{2}\lambda C_{011g} = 0, \\
i\dot{C}_{002g} &=2\Delta_{31} C_{002g} + \sqrt{2}\lambda C_{011g} + \sqrt{2}V e^{i\theta} C_{001e} = 0, \\
i\dot{C}_{110g} &=F C_{010g} + \sqrt{2}\lambda C_{200g} + \Delta_{11} C_{110g} + \Delta_{21} C_{110g} + \lambda C_{101g} + J C_{010e} + \sqrt{2}\lambda C_{020g} = 0, \\
i\dot{C}_{011g} &=\sqrt{2}\lambda C_{020g} + \sqrt{2}\lambda C_{002g} + \lambda C_{101g} + \Delta_{21} C_{011g} + \Delta_{31} C_{011g} + V e^{i\theta} C_{010e} = 0, \\
i\dot{C}_{101g} &=F C_{001g} + \lambda C_{110g} + \Delta_{11} C_{101g} + \Delta_{31} C_{101g} + \lambda C_{011g} + V e^{i\theta} C_{100e} + J C_{001e} = 0,\\
i\dot{C}_{100e} &=F C_{000e} + \sqrt{2}J C_{200g} + V e^{-i\theta} C_{101g} + \Delta_{11} C_{100e} + \Delta_{01} C_{100e} + \lambda C_{010e} = 0, \\
i\dot{C}_{010e} &=J C_{110g} + V e^{-i\theta} C_{011g} + \lambda C_{100e} + \Delta_{21} C_{010e} + \Delta_{01} C_{010e} + \lambda C_{001e} = 0, \\
i\dot{C}_{001e} &=\sqrt{2}V e^{-i\theta} C_{002g} + J C_{101g} + \lambda C_{010e} + \Delta_{31} C_{001e} + \Delta_{01} C_{001e} = 0.
\label{C200g}
\end{aligned}
\end{equation}
Assuming $\Delta_{21}  = \Delta_{11}$ and under the weak driving condition, the probability amplitudes are derived as
\begin{equation}
\begin{aligned}
C_{100g} &= -\frac{F [V^2 \Delta_{11} + \Delta_{01} (-\Delta_{11} \Delta_{31} + \lambda^2)]}{\Delta_{11} (V^2 \Delta_{11} + (J^2 - \Delta_{01} \Delta_{11}) \Delta_{31}) - (J^2 + V^2 - \Delta_{01} (\Delta_{11} + \Delta_{31})) \lambda^2 + 2 J V \lambda^2 \cos(\theta)}, \\
C_{010g} &= -\frac{F (e^{i \theta} J V - V^2 + \Delta_{01} \Delta_{31}) \lambda}{\Delta_{11} (V^2 \Delta_{11} + (J^2 - \Delta_{01} \Delta_{11}) \Delta_{31}) - (J^2 + V^2 - \Delta_{01} (\Delta_{11} + \Delta_{31})) \lambda^2 + 2 J V \lambda^2 \cos(\theta)}, \\
C_{001g} &= \frac{F (e^{i \theta} J V \Delta_{11} + \Delta_{01} \lambda^2)}{\Delta_{11} (V^2 \Delta_{11} + (J^2 - \Delta_{01} \Delta_{11}) \Delta_{31}) - (J^2 + V^2 - \Delta_{01} (\Delta_{11} + \Delta_{31})) \lambda^2 + 2 J V \lambda^2 \cos(\theta)}, \\
C_{200g}&= \frac{A'}{B'},\,\,\,\,\,\,\,\,\,\, C_{020g}= \frac{C'}{B'},\,\,\,\,\,\,\,\,\,\, C_{002g}= \frac{D'}{B'},
\label{A'B'C'D'}
\end{aligned}
\end{equation}
where \(A'\), \(B'\), \(C'\), and \(D'\) are given by
\begin{align}
A'=&F^2 \{ J^2 V^2 \lambda^4 ( 2 V^2 \Delta_{11} - 2 \Delta_{01} \Delta_{11}^2 - 2 \Delta_{11}^3 - 2 \Delta_{01} \Delta_{11} \Delta_{31} - 2 \Delta_{11}^2 \Delta_{31} - \Delta_{01} \Delta_{31}^2 - 2 \Delta_{11} \Delta_{31}^2 - \Delta_{31}^3 + J^2 (\Delta_{11}+ \nonumber\\
 & \Delta_{31})- (3 \Delta_{11} + \Delta_{31}) \lambda^2 ) + e^{3 i \theta} J V \lambda^2 [ V^4 \Delta_{11}^2 (\Delta_{01} + 2 \Delta_{11} + \Delta_{31}) - V^2 (\Delta_{11} \Delta_{31} - \lambda^2) ( \Delta_{11} (2 \Delta_{01} (\Delta_{01} + 2 \Delta_{11}) \nonumber\\
 &+ 3 \Delta_{01} \Delta_{31} + \Delta_{31}^2) - (\Delta_{01} + \Delta_{31}) \lambda^2 ) + (-\Delta_{11} \Delta_{31} + \lambda^2)^2 ( \Delta_{01}^3 + 2 \Delta_{11} (\Delta_{11} + \Delta_{31})^2 + 2 \Delta_{01}^2 (2 \Delta_{11} + \Delta_{31})+ \nonumber\\
& \Delta_{01}  (2 \Delta_{11} + \Delta_{31})^2 + J^2 (\Delta_{01} + 3 \Delta_{11} + 2 \Delta_{31}) - (3 \Delta_{11} + \Delta_{31}) \lambda^2 ) ] + e^{i \theta} J V \lambda^2 \{ V^4 \Delta_{11}^2 (\Delta_{01} + 2 \Delta_{11} + \Delta_{31})  + V^2 \nonumber\\
&( -\Delta_{11}^2 \Delta_{31} (-4 J^2 + 2 \Delta_{01} (\Delta_{01} + 2 \Delta_{11}) + 3 \Delta_{01} \Delta_{31} + \Delta_{31}^2) + 2 \Delta_{11} [-2 J^2 + \Delta_{01}^2 + \Delta_{31}^2 + 2 \Delta_{01} (\Delta_{11} + \Delta_{31})] \lambda^2 \nonumber\\
&+ (\Delta_{01} + 4 \Delta_{11} + \Delta_{31}) \lambda^4 ) + (\Delta_{11} \Delta_{31} - \lambda^2) [ 2 J^4 (\Delta_{11} + \Delta_{31}) + J^2 ( -4 \Delta_{11}^2 (\Delta_{01} + \Delta_{11}) - 5 \Delta_{11} (\Delta_{01} + \Delta_{11}) \Delta_{31} \nonumber\\
&- 2 (\Delta_{01} + 2 \Delta_{11}) \Delta_{31}^2  - 2 \Delta_{31}^3 + (\Delta_{01} - 5 \Delta_{11} - 2 \Delta_{31}) \lambda^2 )  + (\Delta_{11} \Delta_{31} - \lambda^2) ( \Delta_{01}^3 + 2 \Delta_{11} (\Delta_{11} + \Delta_{31})^2 + 2 \Delta_{01}^2 (2 \nonumber\\
& \Delta_{11}+ \Delta_{31}) + \Delta_{01} (2 \Delta_{11} + \Delta_{31})^2 - (3 \Delta_{11} + \Delta_{31}) \lambda^2 ) ] \}+ e^{2 i \theta} [ 2 V^8 \Delta_{11}^3 + J^6 (\Delta_{11} + \Delta_{31}) (-\Delta_{11} \Delta_{31} + \lambda^2)^2+ V^6  \nonumber\\
&\Delta_{11}^2 ( -2 \Delta_{11} (2 \Delta_{11} (\Delta_{01} + \Delta_{11}) + 2 (2 \Delta_{01} + \Delta_{11}) \Delta_{31} + \Delta_{31}^2) + 5 (\Delta_{01} - \Delta_{11}) \lambda^2 ) + \Delta_{01} (-\Delta_{11} \Delta_{31} + \lambda^2)^2 ( (\Delta_{01} \nonumber\\
&+ \Delta_{11})^2 (\Delta_{01} + \Delta_{31}) - (2 \Delta_{01} + \Delta_{11} + \Delta_{31}) \lambda^2 ) ( 2 \Delta_{11} (\Delta_{11} + \Delta_{31})^2 - (3 \Delta_{11} + \Delta_{31}) \lambda^2 ) + V^4 ( 2 \Delta_{11}^3 ( \Delta_{11}^2 (\Delta_{01}+\nonumber\\
&  \Delta_{11})^2 + 2 \Delta_{11} (\Delta_{01} + \Delta_{11}) (3 \Delta_{01} + \Delta_{11}) \Delta_{31} + (6 \Delta_{01}^2 + 8 \Delta_{01} \Delta_{11} + 3 \Delta_{11}^2) \Delta_{31}^2 + (3 \Delta_{01} + 2 \Delta_{11}) \Delta_{31}^3 )- \Delta_{11}^2 ( \Delta_{01}^2 \nonumber\\
& (11 \Delta_{11} + 15 \Delta_{31}) + \Delta_{11} (3 \Delta_{11}^2 + 7 \Delta_{11} \Delta_{31} + 3 \Delta_{31}^2)  + \Delta_{01} (12 \Delta_{11}^2 + 2 \Delta_{11} \Delta_{31} + 5 \Delta_{31}^2) ) \lambda^2  + \Delta_{11} [4 \Delta_{01}^2 - 9 \Delta_{01} \Delta_{11} \nonumber\\
&+ 3 \Delta_{11} (2 \Delta_{11} + \Delta_{31})] \lambda^4 + (\Delta_{01} + \Delta_{31}) \lambda^6 ) - V^2 (\Delta_{11} \Delta_{31} - \lambda^2) ( 2 \Delta_{11}^2 (\Delta_{01} + \Delta_{11}) (\Delta_{11} + \Delta_{31}) ( 2 \Delta_{01} \Delta_{11} (\Delta_{01} + \nonumber\\
& \Delta_{11}) + (4 \Delta_{01}^2 + 3 \Delta_{01} \Delta_{11} + \Delta_{11}^2) \Delta_{31} + (3 \Delta_{01} + \Delta_{11}) \Delta_{31}^2 )  - \Delta_{11} ( 2 \Delta_{11} (5 \Delta_{01}^3 + 7 \Delta_{01}^2 \Delta_{11} + 5 \Delta_{01} \Delta_{11}^2 + \Delta_{11}^3)+ ( \nonumber\\
&7 \Delta_{01}^3 + 8 \Delta_{01}^2 \Delta_{11} + 22 \Delta_{01} \Delta_{11}^2 + 9 \Delta_{11}^3) \Delta_{31}  + (4 \Delta_{01}^2 + 9 \Delta_{01} \Delta_{11} + 7 \Delta_{11}^2) \Delta_{31}^2 + 2 \Delta_{11} \Delta_{31}^3 ) \lambda^2 + [ \Delta_{01}^3 - 6 \Delta_{01}^2 \Delta_{11}\nonumber\\
& + \Delta_{11} (\Delta_{11} - \Delta_{31}) (\Delta_{11} + \Delta_{31}) + \Delta_{01} \Delta_{11} (8 \Delta_{11} + \Delta_{31})] \lambda^4  + (3 \Delta_{11} + \Delta_{31}) \lambda^6 )+ J^4 [ V^4 \Delta_{11}^2 (\Delta_{11} + \Delta_{31}) - 2 V^2  \nonumber\\
&\Delta_{11}(\Delta_{11} \Delta_{31} - \lambda^2) (-\Delta_{11} \Delta_{31} + \Delta_{01} (\Delta_{11} + \Delta_{31}) + \lambda^2) - (-\Delta_{11} \Delta_{31} + \lambda^2)^2 ( (\Delta_{11} + \Delta_{31}) (-\Delta_{01}^2 + 2 \Delta_{01} \Delta_{11} + 2   \nonumber\\
&\Delta_{11}^2+ (\Delta_{01} + \Delta_{11}) \Delta_{31} + \Delta_{31}^2) + (-\Delta_{01} + 2 \Delta_{11} + \Delta_{31}) \lambda^2 ) ] + J^2 ( V^6 \Delta_{11}^2 (3 \Delta_{11} + \Delta_{31}) - V^4 \Delta_{11} ( \Delta_{01} \Delta_{11} (3 \Delta_{11} + \nonumber\\
&\Delta_{31}) (\Delta_{11} + 3 \Delta_{31})  + \Delta_{11} (\Delta_{11} + \Delta_{31}) (3 \Delta_{11}^2 + \Delta_{11} \Delta_{31} + \Delta_{31}^2) + (-7 \Delta_{01} \Delta_{11} + 3 \Delta_{11}^2 - 2 \Delta_{01} \Delta_{31} + 2 \Delta_{11} \Delta_{31}) \lambda^2 )\nonumber\\
 &- (-\Delta_{11} \Delta_{31} + \lambda^2)^2 ( (\Delta_{01} + \Delta_{11}) (\Delta_{11} + \Delta_{31}) [\Delta_{01} \Delta_{31} (-\Delta_{11} + \Delta_{31}) - 2 \Delta_{11} \Delta_{31} (\Delta_{11} + \Delta_{31}) + \Delta_{01}^2 (3 \Delta_{11} + \Delta_{31}\nonumber\\
 &)] + [-\Delta_{01}^3 + 2 \Delta_{01}^2 (2 \Delta_{11} + \Delta_{31}) + \Delta_{11} (2 \Delta_{11} + \Delta_{31}) (\Delta_{11} + 3 \Delta_{31}) + \Delta_{01} (3 \Delta_{11}^2 + 3 \Delta_{11} \Delta_{31} + \Delta_{31}^2)] \lambda^2 - (3 \Delta_{11} \nonumber\\
 &+ \Delta_{31}) \lambda^4 ) + V^2 (\Delta_{11} \Delta_{31} - \lambda^2) ( \Delta_{11} (6 \Delta_{01} \Delta_{11}^2 (\Delta_{01} + \Delta_{11}) + \Delta_{11} (11 \Delta_{01}^2 + 9 \Delta_{01} \Delta_{11} + \Delta_{11}^2) \Delta_{31}  + 3 \Delta_{01} (\Delta_{01}+\nonumber\\
  & \Delta_{11}) \Delta_{31}^2 + (2 \Delta_{01} - \Delta_{11}) \Delta_{31}^3) - [-5 \Delta_{01} \Delta_{11} (\Delta_{11} + \Delta_{31}) + \Delta_{01}^2 (5 \Delta_{11} + \Delta_{31}) + \Delta_{11} (\Delta_{11} - \Delta_{31}) (\Delta_{11} + 2 \Delta_{31})] \nonumber\\
&\lambda^2 + (2 \Delta_{01} + 7 \Delta_{11} + 3 \Delta_{31}) \lambda^4 ) )]\},\nonumber\\
B' = &\sqrt{2} \{J^2 V^2 (\Delta_{01} + 2 \Delta_{11} + \Delta_{31}) \lambda^4+ e^{4 i \theta} J^2 V^2 (\Delta_{01} + 2 \Delta_{11} + \Delta_{31}) \lambda^4 + e^{i \theta} J V \lambda^2 [J^4 (\Delta_{11} + \Delta_{31})+ \Delta_{11} [2 V^4 + 2\nonumber\\
&\Delta_{11}^2 (\Delta_{01} + \Delta_{11})^2+ \Delta_{11} (3 \Delta_{01}^2 + 6 \Delta_{01} \Delta_{11} + 4 \Delta_{11}^2) \Delta_{31} + 2 (\Delta_{01}^2 + 3 \Delta_{01} \Delta_{11} + 3 \Delta_{11}^2) \Delta_{31}^2+ (2 \Delta_{01} + 3 \Delta_{11}) \Delta_{31}^3- \nonumber\\
&V^2(3 \Delta_{01} \Delta_{11} + 2 \Delta_{11}^2 + 4 \Delta_{01} \Delta_{31}+ 3 \Delta_{11} \Delta_{31} + 2 \Delta_{31}^2)] - [2 \Delta_{01}^2 \Delta_{11} + 3 \Delta_{11}^3 + 7 \Delta_{11}^2 \Delta_{31}+ \Delta_{11} \Delta_{31}^2 - \Delta_{31}^3 + \Delta_{01}\nonumber\\
&(\Delta_{11} - \Delta_{31}) (2 \Delta_{11} + \Delta_{31})+ V^2 (7 \Delta_{11} + \Delta_{31})] \lambda^2+ 2 (3 \Delta_{11} + \Delta_{31}) \lambda^4 - J^2 (3 \Delta_{11}^3 + 2 \Delta_{11}^2 \Delta_{31} + \Delta_{11} \Delta_{31}^2 + \Delta_{31}^3- \nonumber\\
&V^2 (3 \Delta_{11} + \Delta_{31}) + \Delta_{01} (3 \Delta_{11}^2 + 3 \Delta_{11} \Delta_{31} + \Delta_{31}^2)+ (5 \Delta_{11} + 3 \Delta_{31}) \lambda^2)] + e^{3 i \theta} J V \lambda^2 [J^4 (\Delta_{11} + \Delta_{31})+ \Delta_{11} [2 V^4\nonumber\\
& + 2 \Delta_{11}^2 (\Delta_{01} + \Delta_{11})^2+ \Delta_{11} (3 \Delta_{01}^2 + 6 \Delta_{01} \Delta_{11} + 4 \Delta_{11}^2) \Delta_{31}+ 2 (\Delta_{01}^2 + 3 \Delta_{01} \Delta_{11} + 3 \Delta_{11}^2) \Delta_{31}^2+ (2 \Delta_{01} + 3 \Delta_{11})\nonumber\\
 &\Delta_{31}^3- V^2 (3 \Delta_{01} \Delta_{11} + 2 \Delta_{11}^2 + 4 \Delta_{01} \Delta_{31}+ 3 \Delta_{11} \Delta_{31} + 2 \Delta_{31}^2)] - [2 \Delta_{01}^2 \Delta_{11} + 3 \Delta_{11}^3 + 7 \Delta_{11}^2 \Delta_{31}+ \Delta_{11} \Delta_{31}^2 - \Delta_{31}^3 \nonumber\\
&+ \Delta_{01} (\Delta_{11} - \Delta_{31}) (2 \Delta_{11} + \Delta_{31})+ V^2 (7 \Delta_{11} + \Delta_{31})] \lambda^2+ 2 (3 \Delta_{11} + \Delta_{31}) \lambda^4 - J^2 (3 \Delta_{11}^3 + 2 \Delta_{11}^2 \Delta_{31} + \Delta_{11} \Delta_{31}^2 \nonumber\\
&+ \Delta_{31}^3- V^2 (3 \Delta_{11} + \Delta_{31}) + \Delta_{01} (3 \Delta_{11}^2 + 3 \Delta_{11} \Delta_{31} + \Delta_{31}^2)+ (5 \Delta_{11} + 3 \Delta_{31}) \lambda^2)] + e^{2 i \theta} [2 V^6 \Delta_{11} (\Delta_{11} - \lambda) (\Delta_{11} \nonumber\\
&+ \lambda)+ J^6 (\Delta_{11} + \Delta_{31}) (\Delta_{11} \Delta_{31} - \lambda^2) - (\Delta_{11}^2 \Delta_{31} - (\Delta_{11} + \Delta_{31}) \lambda^2)((\Delta_{01} + \Delta_{11})^2 (\Delta_{01} + \Delta_{31})- (2 \Delta_{01} + \Delta_{11} \nonumber\\
&+ \Delta_{31}) \lambda^2) (2 \Delta_{11} (\Delta_{11} + \Delta_{31})^2- (3 \Delta_{11} + \Delta_{31}) \lambda^2) + J^4 (V^2 \Delta_{11} (\Delta_{11}^2 + 4 \Delta_{11} \Delta_{31} + \Delta_{31}^2)- \Delta_{11} \Delta_{31} (\Delta_{11}+ \Delta_{31})\nonumber\\
 & (\Delta_{11} (4 \Delta_{01} + 3 \Delta_{11})+ (\Delta_{01} + \Delta_{11}) \Delta_{31} + \Delta_{31}^2) + [\Delta_{11}^2 (4 \Delta_{01} + 3 \Delta_{11}) + \Delta_{11} (7 \Delta_{01} + \Delta_{11}) \Delta_{31}+ 2 \Delta_{01} \Delta_{31}^2 + \Delta_{31}^3 \nonumber\\
 &- 2 V^2 (2 \Delta_{11} + \Delta_{31})] \lambda^2+ (-\Delta_{01} + 3 \Delta_{11} + 2 \Delta_{31}) \lambda^4) + J^2 (\Delta_{11}^2 [-V^2 + (\Delta_{01} + \Delta_{11}) (\Delta_{11} + \Delta_{31})][-3 V^2 (\Delta_{11} + \nonumber\\
 &\Delta_{31}) + \Delta_{31} (\Delta_{11} (5 \Delta_{01} + 2 \Delta_{11})+ 3 (\Delta_{01} + \Delta_{11}) \Delta_{31} + 3 \Delta_{31}^2)] - \{\Delta_{11}^3 (\Delta_{01} + \Delta_{11}) (5 \Delta_{01} + 2 \Delta_{11})+ 3 \Delta_{11}^2 (5 \Delta_{01}^2 \nonumber\\
 &+ 6 \Delta_{01} \Delta_{11} + 3 \Delta_{11}^2) \Delta_{31}+ \Delta_{11} (8 \Delta_{01}^2 + 15 \Delta_{01} \Delta_{11} + 15 \Delta_{11}^2) \Delta_{31}^2 + (\Delta_{01} + \Delta_{11}) (\Delta_{01} + 4 \Delta_{11}) \Delta_{31}^3+ (\Delta_{01} - \Delta_{11})  \nonumber\\
&\Delta_{31}^4+ V^4 (5 \Delta_{11} + \Delta_{31})- V^2 [\Delta_{01} (\Delta_{11} + \Delta_{31}) (11 \Delta_{11} + 2 \Delta_{31})+ (4 \Delta_{11} + \Delta_{31}) (\Delta_{11}^2 + \Delta_{31}^2)]\} \lambda^2 + [4 \Delta_{11}^3 + 15  \nonumber\\
&\Delta_{11}^2\Delta_{31} + 4 \Delta_{11} \Delta_{31}^2 - \Delta_{31}^3+ 2 \Delta_{01}^2 (2 \Delta_{11} + \Delta_{31})
+ \Delta_{01} (\Delta_{11} - \Delta_{31}) (2 \Delta_{11} + \Delta_{31})+ 4 V^2 (3 \Delta_{11} + \Delta_{31})] \lambda^4 - 2 (3 \nonumber\\
& \Delta_{11}+ \Delta_{31}) \lambda^6) + V^2 (2 \Delta_{11}^3 (\Delta_{01} + \Delta_{11}) (\Delta_{11} + \Delta_{31})(\Delta_{11} (\Delta_{01} + \Delta_{11}) + (3 \Delta_{01} + \Delta_{11}) \Delta_{31} + 2 \Delta_{31}^2) - \Delta_{11} (\Delta_{11}^2\nonumber\\
& (9 \Delta_{01}^2 + 12 \Delta_{01} \Delta_{11} + 5 \Delta_{11}^2)+ \Delta_{11} (14 \Delta_{01}^2 + 15 \Delta_{01} \Delta_{11} + 11 \Delta_{11}^2) \Delta_{31}+ 3 (\Delta_{01} + \Delta_{11}) (2 \Delta_{01} + 3 \Delta_{11}) \Delta_{31}^2+ 4 (\Delta_{01} \nonumber\\
&+ \Delta_{11}) \Delta_{31}^3) \lambda^2 + [\Delta_{01} \Delta_{31} (-\Delta_{11} + \Delta_{31}) + 2 \Delta_{01}^2 (2 \Delta_{11} + \Delta_{31})+ \Delta_{11} (9 \Delta_{11}^2 + 10 \Delta_{11} \Delta_{31} + 3 \Delta_{31}^2)] \lambda^4- 2 (3 \Delta_{11} +  \nonumber\\
&\Delta_{31}) \lambda^6)- V^4 (\Delta_{11} (\Delta_{11} - \lambda) (\Delta_{11} + \lambda)(4 \Delta_{11}^2 + 4 \Delta_{11} \Delta_{31} + 2 \Delta_{31}^2 + 5 \lambda^2)+ \Delta_{01} (2 \Delta_{11}^3 (2 \Delta_{11} + 3 \Delta_{31})- \Delta_{11} (7 \Delta_{11}\nonumber\\
 &+ 6 \Delta_{31}) \lambda^2 + \lambda^4)) ] \}  [ \Delta_{11} (V^2 \Delta_{11} + (J^2 - \Delta_{01} \Delta_{11}) \Delta_{31})- [J^2 + V^2 - \Delta_{01} (\Delta_{11} + \Delta_{31})] \lambda^2+ 2 J V \lambda^2 \cos(\theta) ], \nonumber\\
C'=&F^2 \lambda^2 \{e^{5 i \theta}J^3 V^3 (\Delta_{01} + 2 \Delta_{11} + \Delta_{31}) \lambda^2- J^2 V^2 (\Delta_{01} + 2 \Delta_{11} + \Delta_{31}) \lambda^4 + e^{4 i \theta}J^2 V^2 (J^4 (\Delta_{11} + \Delta_{31})+ 2 \Delta_{11} (-V^2\nonumber\\
 &+ \Delta_{11} (\Delta_{01} + \Delta_{11}) + (\Delta_{01} + \Delta_{11}) \Delta_{31} + \Delta_{31}^2)[-V^2 + (\Delta_{01} + \Delta_{11}) (\Delta_{11} + \Delta_{31})] + [\Delta_{01}^2 (-3 \Delta_{11} + \Delta_{31})- V^2 (\nonumber\\
&\Delta_{01} + 9 \Delta_{11} + 2 \Delta_{31})- (\Delta_{11} + \Delta_{31}) (3 \Delta_{11}^2 - \Delta_{31}^2)+ 2 \Delta_{01} (-2 \Delta_{11}^2 + 2 \Delta_{11} \Delta_{31} + \Delta_{31}^2)] \lambda^2 + (-\Delta_{01} + 4 \Delta_{11} + \Delta_{31}\nonumber\\
&) \lambda^4+ J^2 (V^2 (3 \Delta_{11} + \Delta_{31})- (\Delta_{11} + \Delta_{31}) (3 \Delta_{11} (\Delta_{01} + \Delta_{11}) + (\Delta_{01} + \Delta_{11}) \Delta_{31} + \Delta_{31}^2)+ (\Delta_{01} - 3 \Delta_{11} - 2 \Delta_{31})  \nonumber\\
&\lambda^2))+ e^{i \theta}J V \lambda^2 (V^4 (\Delta_{01} + 2 \Delta_{11} + \Delta_{31})+ V^2 (-\Delta_{31} (2 \Delta_{01} (\Delta_{01} + 2 \Delta_{11}) + 3 \Delta_{01} \Delta_{31} + \Delta_{31}^2)- 2 (\Delta_{01} + 2 \Delta_{11} +   \nonumber\\
&\Delta_{31})\lambda^2)+ J^2 (-\Delta_{31} (2 \Delta_{11} (\Delta_{01} + 2 \Delta_{11}) + (\Delta_{01} + 3 \Delta_{11}) \Delta_{31})+ 2 (\Delta_{01} + 2 \Delta_{11} + \Delta_{31}) \lambda^2)+ \Delta_{31}^2 (\Delta_{01}^3 + 2 \Delta_{11} ( \nonumber\\
&\Delta_{11}+ \Delta_{31})^2 + 2 \Delta_{01}^2 (2 \Delta_{11} + \Delta_{31}) + \Delta_{01} (2 \Delta_{11} + \Delta_{31})^2- (3 \Delta_{11} + \Delta_{31}) \lambda^2)) + e^{3 i \theta}J V \{-4 V^6 \Delta_{11}+ 2 J^4 (\Delta_{11} +  \nonumber\\
&\Delta_{31})(-V^2 + \Delta_{01} \Delta_{31}) + V^4 (4 \Delta_{11} (2 \Delta_{11} (\Delta_{01} + \Delta_{11}) + (3 \Delta_{01} + 2 \Delta_{11}) \Delta_{31} + \Delta_{31}^2)+ (-\Delta_{01} + 12 \Delta_{11} + \Delta_{31}) \lambda^2) \nonumber\\
&+ V^2 (-4 \Delta_{11} (\Delta_{01} + \Delta_{11}) (\Delta_{11} + \Delta_{31}) (\Delta_{11} (\Delta_{01} + \Delta_{11})+ (3 \Delta_{01} + \Delta_{11}) \Delta_{31} + 2 \Delta_{31}^2)+ (2 \Delta_{11} (3 \Delta_{01}^2 + 4 \Delta_{01} \Delta_{11}  \nonumber\\
&+3 \Delta_{11}^2)+ 2 (\Delta_{01}^2 - 5 \Delta_{01} \Delta_{11} + 5 \Delta_{11}^2) \Delta_{31}- (\Delta_{01} - 2 \Delta_{11}) \Delta_{31}^2 - \Delta_{31}^3) \lambda^2 + 2 (\Delta_{01} - 4 \Delta_{11} - \Delta_{31}) \lambda^4)+ \Delta_{31} (4 \Delta_{11} \nonumber\\
 &(\Delta_{01}+ \Delta_{11})^2 (\Delta_{01} + \Delta_{31}) (\Delta_{11} + \Delta_{31})^2 - (6 \Delta_{01}^2 \Delta_{11} (2 \Delta_{11} + \Delta_{31}) + \Delta_{01}^3 (6 \Delta_{11} + \Delta_{31})+ \Delta_{01} (14 \Delta_{11}^3 + 26 \Delta_{11}^2 \Delta_{31} \nonumber\\
 &+ 8 \Delta_{11} \Delta_{31}^2 - \Delta_{31}^3)+ 2 \Delta_{11} (2 \Delta_{11}^3 + 8 \Delta_{11}^2 \Delta_{31} + 5 \Delta_{11} \Delta_{31}^2 + \Delta_{31}^3)) \lambda^2 + (4 \Delta_{01} + 2 \Delta_{11} + \Delta_{31}) (3 \Delta_{11} + \Delta_{31}) \lambda^4)- \nonumber\\
& J^2 [2 V^4 (3 \Delta_{11} + \Delta_{31})+ 2 \Delta_{31} (3 \Delta_{01} \Delta_{11}^2 (\Delta_{01} + \Delta_{11}) + \Delta_{11} (4 \Delta_{01}^2 + 6 \Delta_{01} \Delta_{11} + 3 \Delta_{11}^2) \Delta_{31}+ (\Delta_{01} + \Delta_{11})^2 \Delta_{31}^2 +  \nonumber\\
&\Delta_{01} \Delta_{31}^3)+ \Delta_{31} [-2 \Delta_{01}^2 + 3 \Delta_{01} \Delta_{31} - 5 \Delta_{11} (2 \Delta_{11} + \Delta_{31})] \lambda^2
+ 2 (\Delta_{01} + 2 \Delta_{11} + \Delta_{31}) \lambda^4 + V^2 (-2 \Delta_{01} (3 \Delta_{11} + \Delta_{31}\nonumber\\
&) (\Delta_{11} + 2 \Delta_{31})- 2 (\Delta_{11} + \Delta_{31}) (3 \Delta_{11}^2 + \Delta_{11} \Delta_{31} + \Delta_{31}^2)+ (\Delta_{01} - 8 \Delta_{11} - 5 \Delta_{31}) \lambda^2)]\} + e^{2 i \theta} [2 V^8 \Delta_{11}+ V^6 (-2 \Delta_{11} \nonumber\\
&(2 \Delta_{11} (\Delta_{01} + \Delta_{11}) + 2 (2 \Delta_{01} + \Delta_{11}) \Delta_{31} + \Delta_{31}^2)+ (\Delta_{01} - 5 \Delta_{11}) \lambda^2)+ \Delta_{01} \Delta_{31}^2 ((\Delta_{01} + \Delta_{11})^2 (\Delta_{01} + \Delta_{31}) - (2 \nonumber\\
&\Delta_{01} + \Delta_{11} + \Delta_{31}) \lambda^2)(2 \Delta_{11} (\Delta_{11} + \Delta_{31})^2 - (3 \Delta_{11} + \Delta_{31}) \lambda^2)+ J^4 ((\Delta_{11} + \Delta_{31}) (V^4 - 2 V^2 \Delta_{01} \Delta_{31} + (\Delta_{01} - 2 \Delta_{11}\nonumber\\
 &)(\Delta_{01} + \Delta_{11}) \Delta_{31}^2)+ \Delta_{31} (2 \Delta_{11} (\Delta_{01} + 2 \Delta_{11}) + (\Delta_{01} + 3 \Delta_{11}) \Delta_{31}) \lambda^2 - (\Delta_{01} + 2 \Delta_{11}+ \Delta_{31}) \lambda^4) + V^2 \Delta_{31} (-2 \Delta_{11} (\nonumber\\
&\Delta_{01}  + \Delta_{11}) (\Delta_{11} + \Delta_{31}) (2 \Delta_{01} \Delta_{11} (\Delta_{01} + \Delta_{11})+ (4 \Delta_{01}^2 + 3 \Delta_{01} \Delta_{11} + \Delta_{11}^2) \Delta_{31}+ (3 \Delta_{01} + \Delta_{11}) \Delta_{31}^2)+ [3 \Delta_{01}^3 (2  \nonumber\\
&\Delta_{11} + \Delta_{31}) + 2 \Delta_{01}^2 (6 \Delta_{11}^2 + 2 \Delta_{11} \Delta_{31} + \Delta_{31}^2)
+ \Delta_{01} \Delta_{11} (14 \Delta_{11}^2 + 26 \Delta_{11} \Delta_{31} + 9 \Delta_{31}^2)+ \Delta_{11} (4 \Delta_{11}^3 + 13 \Delta_{11}^2 \Delta_{31} + \nonumber\\
&9 \Delta_{11} \Delta_{31}^2 + 2 \Delta_{31}^3)] \lambda^2 - (4 \Delta_{01} + 2 \Delta_{11} + \Delta_{31}) (3 \Delta_{11} + \Delta_{31}) \lambda^4) + V^4 (2 \Delta_{11} (\Delta_{11}^2 (\Delta_{01} + \Delta_{11})^2 + 2 \Delta_{11} (\Delta_{01} + \Delta_{11}) \nonumber\\
&(3 \Delta_{01} + \Delta_{11}) \Delta_{31}+ (6 \Delta_{01}^2 + 8 \Delta_{01} \Delta_{11} + 3 \Delta_{11}^2) \Delta_{31}^2+ (3 \Delta_{01} + 2 \Delta_{11}) \Delta_{31}^3) - [3 \Delta_{01}^2 (\Delta_{11} + \Delta_{31}) + \Delta_{01} (4 \Delta_{11}^2 - 6 \nonumber\\
&\Delta_{11} \Delta_{31} + \Delta_{31}^2)+ \Delta_{11} (3 \Delta_{11}^2 + 7 \Delta_{11} \Delta_{31} + 3 \Delta_{31}^2)] \lambda^2+ (-\Delta_{01} + 4 \Delta_{11} + \Delta_{31}) \lambda^4) + J^2 (V^6 (3 \Delta_{11} + \Delta_{31})+ V^2 (\Delta_{31}  \nonumber\\
&(6\Delta_{01} \Delta_{11}^2 (\Delta_{01} + \Delta_{11}) + \Delta_{11} (11 \Delta_{01}^2 + 9 \Delta_{01} \Delta_{11} + \Delta_{11}^2) \Delta_{31}+ 3 \Delta_{01} (\Delta_{01} + \Delta_{11}) \Delta_{31}^2 + (2 \Delta_{01} - \Delta_{11}) \Delta_{31}^3)+ \Delta_{31} ( \nonumber\\
&4 \Delta_{01} \Delta_{11} - 10 \Delta_{11}^2 + 6 \Delta_{01} \Delta_{31} - 5 \Delta_{11} \Delta_{31} + \Delta_{31}^2) \lambda^2
+ 4 (\Delta_{01} + 2 \Delta_{11} + \Delta_{31}) \lambda^4) + \Delta_{31}^2 ((\Delta_{01} + \Delta_{11}) (\Delta_{11} + \Delta_{31})\nonumber\\
& [\Delta_{01} (\Delta_{11} - \Delta_{31}) \Delta_{31}+ 2 \Delta_{11} \Delta_{31} (\Delta_{11} + \Delta_{31}) - \Delta_{01}^2 (3 \Delta_{11} + \Delta_{31})]- [-\Delta_{01}^3 + 2 \Delta_{01}^2 (2 \Delta_{11} + \Delta_{31})+ \Delta_{11} (2 \Delta_{11} +  \nonumber\\
&\Delta_{31}) (\Delta_{11} + 3 \Delta_{31}) + \Delta_{01} (3 \Delta_{11}^2 + 3 \Delta_{11} \Delta_{31} + \Delta_{31}^2)] \lambda^2
+ (3 \Delta_{11} + \Delta_{31}) \lambda^4) - V^4 (3 \Delta_{11}^3 + 4 \Delta_{11}^2 \Delta_{31} + 2 \Delta_{11} \Delta_{31}^2 \nonumber\\
&+ \Delta_{31}^3+ (7 \Delta_{11} + 4 \Delta_{31}) \lambda^2 + \Delta_{01} ((3 \Delta_{11} + \Delta_{31}) (\Delta_{11} + 3 \Delta_{31}) + \lambda^2)))]\},\nonumber\\
D' = &e^{i \theta}F^2 \{e^{4 i \theta}J^3 V^3 \Delta_{11}^2 (\Delta_{01} + 2 \Delta_{11} + \Delta_{31}) \lambda^2 + J V \lambda^6 (\Delta_{01}^3 + 4 \Delta_{01}^2 \Delta_{11} + 4 \Delta_{01} \Delta_{11}^2 + 2 \Delta_{11}^3- J^2 (\Delta_{01} + \Delta_{11}) + 2  \nonumber\\
&\Delta_{01}^2 \Delta_{31}+ 4 \Delta_{01} \Delta_{11} \Delta_{31} + 4 \Delta_{11}^2 \Delta_{31} + \Delta_{01} \Delta_{31}^2 + 2 \Delta_{11} \Delta_{31}^2+ V^2 (\Delta_{01} + 4 \Delta_{11} + \Delta_{31}) - (3 \Delta_{11} + \Delta_{31}) \lambda^2)+ e^{3 i \theta}\nonumber\\
&J^2 V^2 [ J^4 \Delta_{11}^2 (\Delta_{11} + \Delta_{31})+ 2 \Delta_{11}^3 [V^2 - (\Delta_{01} + \Delta_{11}) (\Delta_{11} + \Delta_{31})]^2 + \Delta_{11}^2 [V^2 (\Delta_{01} - 5 \Delta_{11} - 4 \Delta_{31}) - \Delta_{01}^2 (3 \Delta_{11} \nonumber\\
&+ \Delta_{31})+ 4 \Delta_{01} (-\Delta_{11}^2 + \Delta_{11} \Delta_{31} + \Delta_{31}^2)+ \Delta_{11} (-3 \Delta_{11}^2 + \Delta_{11} \Delta_{31} + 2 \Delta_{31}^2)] \lambda^2 + \Delta_{11} [2 V^2 + 2 \Delta_{01}^2 + \Delta_{01} \Delta_{11} + (\nonumber\\
 &2 \Delta_{11} + \Delta_{31})(\Delta_{11} + 2 \Delta_{31})] \lambda^4- (3 \Delta_{11} + \Delta_{31}) \lambda^6 + J^2 (-\Delta_{11}^2 (3 \Delta_{11} + \Delta_{31}) [-V^2 + (\Delta_{01} + \Delta_{11}) (\Delta_{11} + \Delta_{31})]+ \nonumber\\
 &\Delta_{11} ((\Delta_{01} - 3 \Delta_{11}) \Delta_{11} - 4 \Delta_{11} \Delta_{31} - 2 \Delta_{31}^2) \lambda^2+ (\Delta_{11} + \Delta_{31}) \lambda^4)] + e^{2 i \theta}J V \lambda^2 \{2 J^4 \Delta_{01} \Delta_{11} (\Delta_{11} + \Delta_{31})+ 4 \Delta_{11}^2 \nonumber\\
 &(\Delta_{01} + \Delta_{11})^2 (\Delta_{01} + \Delta_{31}) (\Delta_{11} + \Delta_{31})^2 - 2 \Delta_{11} [\Delta_{01}^3 (3 \Delta_{11} + \Delta_{31}) + \Delta_{01}^2 (2 \Delta_{11} + \Delta_{31}) (3 \Delta_{11} + \Delta_{31})+ \Delta_{01} \Delta_{11}\nonumber\\
  &(7 \Delta_{11}^2 + 15 \Delta_{11} \Delta_{31} + 6 \Delta_{31}^2)+ \Delta_{11} (2 \Delta_{11}^3 + 9 \Delta_{11}^2 \Delta_{31} + 7 \Delta_{11} \Delta_{31}^2 + 2 \Delta_{31}^3)] \lambda^2 + (\Delta_{01}^3 + 2 \Delta_{01}^2 (2 \Delta_{11} + \Delta_{31})+ 4 \Delta_{11}\nonumber\\
& (\Delta_{11} + \Delta_{31}) (2 \Delta_{11} + \Delta_{31})+ \Delta_{01} (4 \Delta_{11} + \Delta_{31})^2) \lambda^4- (3 \Delta_{11} + \Delta_{31}) \lambda^6 + 4 V^4 \Delta_{11} (\Delta_{11} (\Delta_{01} + \Delta_{31}) - \lambda^2)+ V^2 (-8\nonumber\\
 &\Delta_{11}^2 (\Delta_{01} + \Delta_{11}) (\Delta_{01} + \Delta_{31}) (\Delta_{11} + \Delta_{31}) + 2 \Delta_{11} [\Delta_{01}^2 - \Delta_{01} (\Delta_{11} - 3 \Delta_{31}) + \Delta_{11} (4 \Delta_{11} + \Delta_{31})] \lambda^2+ (-\Delta_{01} + 6 \nonumber\\
  &\Delta_{11}+ \Delta_{31}) \lambda^4) + J^2 [2 \Delta_{11} (-3 \Delta_{01} \Delta_{11}^2 (\Delta_{01} + \Delta_{11}) - \Delta_{11} (\Delta_{01} + \Delta_{11}) (4 \Delta_{01} + \Delta_{11}) \Delta_{31}+ (-\Delta_{01} + \Delta_{11}) (\Delta_{01} + 2 \nonumber\\
  &\Delta_{11}) \Delta_{31}^2 + (-\Delta_{01} + \Delta_{11}) \Delta_{31}^3) + 2 \Delta_{11} ((\Delta_{01} - \Delta_{11})^2 - 3 (\Delta_{01} + \Delta_{11}) \Delta_{31} - 2 \Delta_{31}^2) \lambda^2+ (\Delta_{01} + 3 \Delta_{11} + 2 \Delta_{31}) \lambda^4 +\nonumber\\
& V^2 (\Delta_{11} (7 \Delta_{01} \Delta_{11} + 2 \Delta_{11}^2 + 2 \Delta_{01} \Delta_{31} + 3 \Delta_{11} \Delta_{31} + 2 \Delta_{31}^2)
- 2 (\Delta_{11} + \Delta_{31}) \lambda^2)]\} + e^{i \theta} \lambda^4 [2 V^6 \Delta_{11}+ V^4 (-2 \Delta_{11} [-\nonumber\\
&\Delta_{01}^2 + 2 \Delta_{11} (\Delta_{11} + \Delta_{31}) + \Delta_{01} (2 \Delta_{11} + \Delta_{31})]+ (\Delta_{01} - 3 \Delta_{11}) \lambda^2) + \Delta_{01} ((\Delta_{01} + \Delta_{11})^2 (\Delta_{01} + \Delta_{31})- (2 \Delta_{01} + \Delta_{11}\nonumber\\
& + \Delta_{31}) \lambda^2)(2 \Delta_{11} (\Delta_{11} + \Delta_{31})^2 - (3 \Delta_{11} + \Delta_{31}) \lambda^2) + V^2 (2 \Delta_{11} (-\Delta_{01} + \Delta_{11}) (\Delta_{01} + \Delta_{11}) (\Delta_{11} + \Delta_{31}) (2 \Delta_{01} + \Delta_{11} \nonumber\\
&+ \Delta_{31})- [-\Delta_{01}^3 + 6 \Delta_{01}^2 \Delta_{11} + \Delta_{01} \Delta_{11} (4 \Delta_{11} + 3 \Delta_{31})+ \Delta_{11} (5 \Delta_{11}^2 + 5 \Delta_{11} \Delta_{31} + 2 \Delta_{31}^2)] \lambda^2 + (3 \Delta_{11} + \Delta_{31}) \lambda^4) + \nonumber\\
&J^4 [\Delta_{01}^2 (\Delta_{11} + \Delta_{31}) + \Delta_{01} (-\Delta_{11} \Delta_{31} + \lambda^2) + \Delta_{11} (-\Delta_{11} \Delta_{31} + \lambda^2)] + J^2 (V^4 (\Delta_{11} + \Delta_{31})- (\Delta_{01} + \Delta_{11}) (\Delta_{11} +\nonumber\\
& \Delta_{31}) (\Delta_{01} \Delta_{31} (-\Delta_{11} + \Delta_{31})- 2 \Delta_{11} \Delta_{31} (\Delta_{11}+ \Delta_{31}) + \Delta_{01}^2 (3 \Delta_{11} +\Delta_{31}))+ [\Delta_{01}^3 - 2 \Delta_{01}^2 (2 \Delta_{11} + \Delta_{31}) - \Delta_{11} (2\nonumber\\
& \Delta_{11} + \Delta_{31}) (\Delta_{11} + 3 \Delta_{31})- \Delta_{01} (3 \Delta_{11}^2 + 3 \Delta_{11} \Delta_{31} + \Delta_{31}^2)] \lambda^2+ (3 \Delta_{11} + \Delta_{31}) \lambda^4+ V^2 (5 \Delta_{01}^2 \Delta_{11} + 3 \Delta_{01} \Delta_{11}^2 - \Delta_{11}^3\nonumber\\
& + \Delta_{01}^2 \Delta_{31}+ 5 \Delta_{01} \Delta_{11} \Delta_{31} + 8 \Delta_{11}^2 \Delta_{31} + 3 \Delta_{11} \Delta_{31}^2- (2 \Delta_{01} + 7 \Delta_{11} + 3 \Delta_{31}) \lambda^2))]\},
\end{align}
with $\Delta_{11}=\Delta- \frac{i}{2}\kappa$, $\Delta_{31}=\Delta_3- \frac{i}{2}\kappa$, and $\Delta_{01}=\Delta_0 - \frac{i}{2}\kappa$.
\end{widetext}

\end{document}